\documentclass[twocolumn,secnumarabic,amssymb, nobibnotes, aps, prd]{revtex4-2}

\usepackage{bm}
\usepackage{mathtools,amsmath}
\usepackage{adjustbox}
\usepackage{array,booktabs,tabularx}
\usepackage{siunitx}
\usepackage{upquote}
\usepackage{hhline}
\usepackage{stackengine}
\usepackage{xcolor}
\usepackage{setspace} 
\usepackage{hyperref}

\newcommand{\magn}{magnetization}
\newcommand{\ml}{monolayer~}
\newcommand{\bl}{bilayer~}
\newcommand{\fp}{\textit{ab}~\textit{initio}~calculations}
\newcommand{\data}{\textit{ab}~\textit{initio}~data}
\newcommand{\xyp}{\textit{xy}-plane}

\newcommand{\xzp}{\textit{xz}-plane}

\newcolumntype{C}{>{\centering\arraybackslash}X}

\begin{document}

\begin{abstract}

	The control  of  spin current  is pivotal for spintronic applications,
	especially  for   spin-orbit torque devices.
	Spin Hall effect (SHE) is
	a prevalent method to generate   spin current.
	However,
	it is difficult to manipulate its  spin polarization in nonmagnet.
	Recently, the  discovery of spin current  in ferromagnet
	offers  opportunity to realize the manipulation.
	In the present work,
	the spin current   in van der Waals ferromagnet  Fe$_3$GeTe$_2$  (FGT)
	with varying  magnetization  is  theoretically investigated.
	It has been  observed that the
	spin current in FGT presents
	a nonlinear behavior  with respect to  magnetization.
	The  in-plane and  out-of-plane spin polarizations   emerge  simultaneously, and the bilayer FGT
	can even exhibit arbitrary  polarization
	thanks to the reduced symmetry.
	More intriguingly, the correlation
	between anomalous Hall effect (AHE) and
	spin anomalous Hall effect (SAHE) 
	has been interpreted
	from the aspect of Berry curvature and spin.
	This work illustrates that the interplay of magnetism  and  symmetry
	can effectively control
	the magnitude and polarization of
	the spin current,
	providing a practical method to realize  exotic spin-orbit torques.

\end{abstract}

\title{Controllable Spin Current  in van der Waals Ferromagnet Fe$_3$GeTe$_2$}%

\author{Jiaqi Zhou}
\email{jiaqi.zhou@uclouvain.be}
\affiliation{Institute of Condensed Matter and Nanosciences (IMCN), Université catholique de Louvain, B-1348, Louvain-la-Neuve, Belgium}

\author{Jean-Christophe Charlier}
\email{jean-christophe.charlier@uclouvain.be}
\affiliation{Institute of Condensed Matter and Nanosciences (IMCN), Université catholique de Louvain, B-1348, Louvain-la-Neuve, Belgium}
\date{\today}%
\maketitle

\textit{Introduction.}---The spin-orbit-coupling (SOC) effects
are  the  frontier of spintronics
and  attract    widespread interests.
Versatile  SOC effects  stimulate  fascinating    phenomena,
and   improve   the
performances  of magnetic random access memory (MRAM) \cite{Dieny2020Aug, aelm.201900134}.
For instance,
the perpendicular magnetic anisotropy (PMA) greatly increases
the storage density of MRAM \cite{Ikeda2010Sep},
and the spin-orbit torque (SOT) can
realize ultrafast switching
as well as  low-power writing operation \cite{Ramaswamy2018Sep}.
The last  decade has witnessed the endeavor of
researchers
to realize the  SOT switching \cite{RevModPhys.91.035004}.
The conventional  SOT device  consists of a  ferromagnet/heavy metal bilayer structure,
and the spin current in heavy metal is employed  to
induce SOT and 		switch
the magnetization of ferromagnet \cite{PhysRevLett.109.096602,PhysRevLett.122.077201}.
Spin current can be generated  by
the spin Hall effect (SHE) \cite{RevModPhys.87.1213}.
However,
due to the constraint on SHE imposed by crystal symmetry,
the torque  of
perpendicular-flowing  spin current
is  limited to be in-plane \cite{PhysRevLett.109.096602}.
Thus, the  switching of perpendicular magnetization
requires  an
external magnetic field,
which hinders
the device minimization.
To solve this difficulty,
various methods have been proposed,
such as  the
assistance of  spin-transfer torque \cite{Wang2018Nov},
the asymmetric structural  design  \cite{Yu2014Jul},
as well as  using
low-symmetry crystals as  torque sources \cite{Liu2021Jan,MacNeill2017Mar,PhysRevLett.125.256603}.
More intriguingly, recent efforts have been expanded to
explore the spin current
in ferromagnet  \cite{PhysRevApplied.3.044001,Koike2020Aug,PhysRevLett.125.267204,PhysRevB.99.104414}.
Superior to heavy metal,
the \magn~in ferromagnet  can  break  the  constraint of   symmetry,
thus 	enabling diverse   polarizations  of  spin current.

The  spin current   in ferromagnet 	can be  attributed  to  the
anomalous Hall effect (AHE) \cite{RevModPhys.82.1539}.
When the anomalous Hall current, i.e. spin-polarized current, occurs  in ferromagnet,
it is  naturally accompanied by a pure spin current.
This phenomenon is called  the  spin anomalous Hall effect  (SAHE),
where the spin polarization is collinear with magnetization \cite{Koike2020Aug}.
\textcolor{black}{
	Several experimental and theoretical works have reported the spin current in ferromagnets.
	Generally,
	experimental works claim that
	the spin currents   in  ferromagnets
	completely originate  from  SAHE
	\cite{PhysRevB.99.014403,Koike2020Aug,Iihama2018Feb}.
	However, 
	contradictory  results  appeared regarding 
	whether the spin current is  dependent on the magnetization  of ferromagnet
	\cite{PhysRevB.99.014403,Koike2020Aug,Iihama2018Feb,PhysRevB.100.144427,PhysRevLett.125.267204,PhysRevB.99.104414,PhysRevResearch.2.032053}.
	Meanwhile,  theoretical works also present different interpretations.
	It  has been stated that
	the spin component transverse to the magnetization rapidly precesses and dephases  in a ferromagnet,
	suggesting that 
	the spin current   
	only has the spin along the magnetization,
	and its origin is SAHE exclusively \cite{PhysRevApplied.3.044001}.
	However,
	recent first-principles simulations demonstrate that
	the spin current  in   3\textit{d} ferromagnet  
	is composed of 	two contributions,
	SAHE and SHE.
	The former is dependent on the magnetization
	while the latter is not.
	Moreover, the spin current  with spin component transverse to the magnetization
	is protected from dephasing \cite{PhysRevB.99.220405, DAVIDSON2020126228}.
	Recently, the magnetization-dependent SHE is also reported in 3\textit{d} ferromagnet \cite{PhysRevB.102.144440}.
	In fact, the  debate  is focusing on   
	the different definitions of  SAHE and SHE.
	Indeed, although the previous theoretical works have already 
	provided intuitive understandings of the spin current in ferromagnet,
	rigorous derivations of SAHE 
	based on the linear response theory is still missing.}
Besides,
even if   current studies are
limited to conventional  3\textit{d} ferromagnet,
the development  of
van der Waals (vdW) material
provides  two-dimensional (2D) ferromagnet \cite{Burch2018Nov,Lin2019Jul}
as an  ideal platform to investigate spin current.
The representative  material is the metallic Fe$_3$GeTe$_2$ (FGT),
which can exhibit
gate-tunable room-temperature ferromagnetism \cite{Fei2018Sep,Deng2018Nov}.
Different from other cubic ferromagnets,
FGT is  a uniaxial crystal with low symmetry.
More interestingly, changing the thickness
can  modify the phase of FGT  few-layer,
giving  a broad  space to  investigate
diverse spin current tensors.

In this Letter,
the spin current   in  Fe$_3$GeTe$_2$  ferromagnet
has  been  systematically investigated using \fp.
The monolayer and bilayer FGT  can
exhibit nonlinear spin current  with respect to magnetization,
making  its  magnitude  controllable.
More intriguingly,
both the magnetization  and  the low-symmetry structure    of FGT
can break the constraint  on spin current tensors,
enabling the simultaneous occurrence of
in-plane and out-of-plane  spin  polarizations.
\textcolor{black}{
	Using the Kubo formula,
	the origin of spin current is classified according to the spin,
	and the   correlation  between  AHE and SAHE  is  clarified.}
This work demonstrates that the interplay of magnetism  and  symmetry
can  create  unconventional spin current,
as well as
interprets the origin of spin current in ferromagnet.

\begin{figure}[t]
	\includegraphics[scale=1]{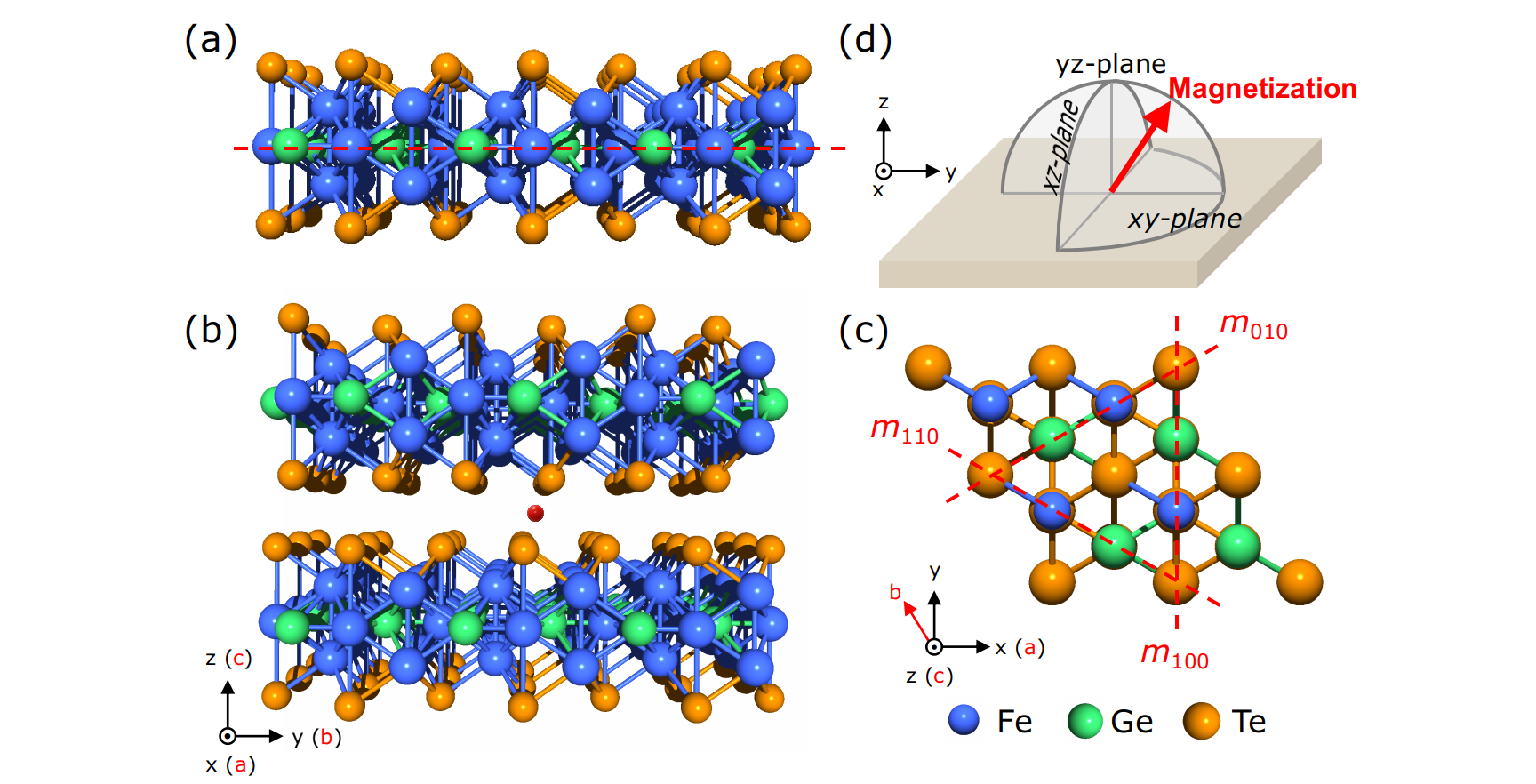}
	\caption{\label{fig:structure}
		(a) Atomic model of \ml FGT with a mirror symmetry (the red dashed line).
		(b) Atomic model of \bl FGT with an inversion center (the red point).
		(c) Top view of \bl FGT including mirror symmetries. Fe, Ge, and Te are represented by blue, green, and orange balls, respectively.
		(d) Illustration of the  magnetization evolution in different planes.
	}
\end{figure}

\textit{Models and methods.}---The
shape of linear response tensors  of an electric field is restricted by the symmetry of crystal \cite{Newnham2005Jan, Gallegolk5043,PhysRevB.92.155138,Freimuth2021Mar}.
Consequently,  crystals in different phases would
exhibit various  Hall currents \cite{PhysRevLett.105.246602}.
FGT is a uniaxial crystal  in the hexagonal structure.
Even if  both   exhibit three mirrors orthogonal to the \xyp~as shown in Fig.~\ref{fig:structure}(c),
the \ml FGT crystallizes in \textit{P}$\bar{6}$\textit{m}2 phase
with a mirror $ \mathcal{M} $ normal to \textit{z}-direction
[see Fig.~\ref{fig:structure}(a)],
while
the \bl FGT is in \textit{P}$\bar{3}$\textit{m}1 phase
with  the spatial inversion $ \mathcal{P} $
[see Fig.~\ref{fig:structure}(b)].
The structural discrepancy creates   opportunity
to explore various Hall currents in these two types of FGT.
The  intrinsic  anomalous  Hall conductivity (AHC)
and  spin Hall  conductivity (SHC) have been investigated using \fp,
as implemented in {\sc Quantum ESPRESSO}  \cite{Giannozzi_2017} and {\sc Wannier90} \cite{Pizzi_2020},
see   details in Supplemental Material \cite{supp}.

\textit{\textcolor{black}{Definitions of Hall conductivities.}}---The
present work focuses  on the current along \textit{x}-axis
when  the  electric field is applied along \textit{y}-axis.
The spin polarization   $\gamma $  is  projected  onto  the
\textit{x}, \textit{y}, or \textit{z}-axis.

In  the 2D system, the  anomalous  Hall conductivity
can  be evaluated  by  Kubo formula as \cite{PhysRevLett.92.037204,PhysRevB.74.195118}
\begin{equation}
	\label{eq:kubo_ahc}
	\sigma_\text{AH}  =
	-\frac{e^2}{\hbar}
	\int_\text{BZ}
	\frac{d^2{k}}{(2\pi)^2}
	\Omega (\bm{k}),
\end{equation}
where
BZ denotes the first Brillouin zone,
$ \Omega (\bm{k}) $ is the Berry curvature,
as
\begin{align}
	\begin{split}
		\label{eq:berry}
		\Omega   (\bm{k}) = \sum_{n}    f_{n\bm{k}}    {\Omega}_{n} (\bm{k}),
	\end{split}
\end{align}
\begin{align}
	\begin{split}
		\label{eq:berry-band}
		\Omega_{n}(\bm{k}) = {\hbar}^2 \sum_{
			m\ne n}\frac{-2\operatorname{Im}[\langle n\bm{k}|
				\hat{v}_x |m\bm{k}\rangle
				\langle m\bm{k}| \hat{v}_{y}|n\bm{k}\rangle]}
		{(\epsilon_{n\bm{k}}-\epsilon_{m\bm{k}})^2},
	\end{split}
\end{align}
where
$  f_{n\bm{k}}  $ is the Fermi-Dirac distribution function,
$ n $ and $ m $ are the band indexes,
$ \epsilon_{n\bm{k}} $ and $ \epsilon_{m\bm{k}} $ are the eigenvalues,
$ \hat{v}_x $  and $ \hat{v}_y $ denote the velocity operators.

In analogy with that of nonmagnet \cite{PhysRevB.98.214402,PhysRevB.99.060408},
the \textcolor{black}{spin Hall  conductivity}   of 2D ferromagnet is given as
\begin{equation}
	\label{eq:kubo_shc}
	\sigma_{\text{SH},\gamma} =
	-\frac{e^2}{\hbar}
	\int_\text{BZ}
	\frac{d^2 {k}}{(2\pi)^2}
	\Omega_{\gamma}(\bm{k}),
\end{equation}
where $ \Omega_\gamma(\bm{k}) $ is
the  \textcolor{black}{spin Berry curvature}, as
\begin{equation}
	\label{eq:berrylike}
	\Omega_\gamma(\bm{k}) = \sum_{n}
	f_{n\bm{k}}  \Omega_{n, \gamma}(\bm{k}),
\end{equation}
\begin{align}
	\begin{split}
		\label{eq:berrylike-band}
		\Omega_{n, \gamma}(\bm{k}) = {\hbar}^2 \sum_{
			m\ne n}\frac{-2\operatorname{Im}[\langle n\bm{k}|
				\frac{1}{2} \{\hat{\sigma}_\gamma, \hat{v}_x\}
				|m\bm{k}\rangle
				\langle m\bm{k}| \hat{v}_{y}|n\bm{k}\rangle]}
		{(\epsilon_{n\bm{k}}-\epsilon_{m\bm{k}})^2},
	\end{split}
\end{align}
where  $ \hat{\sigma}_\gamma $ is the spin operator,
\textcolor{black}{and
	$ \frac{1}{2}\{\hat{\sigma}_\gamma, \hat{v}_x\} =
		\frac{1}{2}(\hat{\sigma}_\gamma \hat{v}_x + \hat{v}_x \hat{\sigma}_\gamma) $.}

Since both AHC  and  SHC  occur in a ferromagnet,
the correlation  between them is worth investigation.
\textcolor{black}{
	Firstly we define a spin matrix $ \bf{S} $
	whose dimension
	is the number of eigenstates,
	and element at the \textit{m}th row and \textit{n}th column  is
	$ \langle m\bm{k}| \hat{\sigma}_\gamma| n\bm{k} \rangle $.
	It can be seen that the \textit{n}th diagonal element of  $ \bf{S} $
	is the spin of the \textit{n}th eigenstate,
	called the \textit{intra} band spin.
	Then,
	through  inserting  a  projection  operator
	$\sum_{t}|t\bm{k}\rangle{\langle}t\bm{k}|~=~1$
	into Eq.~(\ref{eq:berrylike-band})
	and
	considering $ t=n, m,$ respectively,
	the diagonal elements of $ \bf{S} $,
	i.e. $ \langle n\bm{k}| \hat{\sigma}_\gamma| n\bm{k} \rangle$
	and
	$ \langle m\bm{k}| \hat{\sigma}_\gamma | m\bm{k} \rangle $,
	can be extracted.}
Since  both
$   \langle n\bm{k}| \hat{\sigma}_\gamma    | n\bm{k}   \rangle$
and
$ \langle m\bm{k}| \hat{\sigma}_\gamma    | m\bm{k}   \rangle	 $
are real,
this  diagonal part
can be  denoted by $ \bar \Omega_{n, \gamma}(\bm{k}) $  as
\begin{align}
	\begin{split}
		\label{eq:spinberry-band}
		& \bar \Omega_{n, \gamma}(\bm{k}) =  \;
		{\hbar}^2 \sum_{	m\ne n}
		[\langle n\bm{k}| \hat{\sigma}_\gamma    | n\bm{k}   \rangle
			+
			\langle m\bm{k}| \hat{\sigma}_\gamma    | m\bm{k}   \rangle	] \\
		& \hspace{2.2cm}\times \frac{-\operatorname{Im}[
				\langle  n\bm{k}| 	\hat{v}_x    	|m\bm{k}\rangle
				\langle m\bm{k}| \hat{v}_{y}|n\bm{k}\rangle]}
		{(\epsilon_{n\bm{k}}-\epsilon_{m\bm{k}})^2}. \\
	\end{split}
\end{align}
\textcolor{black}{As a result of the \textit{intra} band spin},
Eq.~(\ref{eq:spinberry-band}) is called the
band-resolved \textcolor{black}{ \textit{intra} spin Berry curvature}.
With this quantity and
its summation over the occupied bands,
the  \textcolor{black}{\textit{intra} spin  Hall conductivity}
is  defined  as
	\begin{equation}
		\label{eq:spinberry}
		\bar{\Omega}_{\gamma} (\bm{k})
		= \sum_{n}  f_{n\bm{k}}  \bar{\Omega}_{n, \gamma}(\bm{k}),
	\end{equation}
	\begin{equation}
		\label{eq:kubo_sahc}
		\sigma_{\text{SH},\gamma}^{intra}   =
		-\frac{e^2}{\hbar}
		\int_\text{BZ}
		\frac{d^2 {k}}{(2\pi)^2}
		\bar\Omega_{\gamma}  (\bm{k}).
	\end{equation}

\begin{figure}[b!]
	\includegraphics[scale=1]{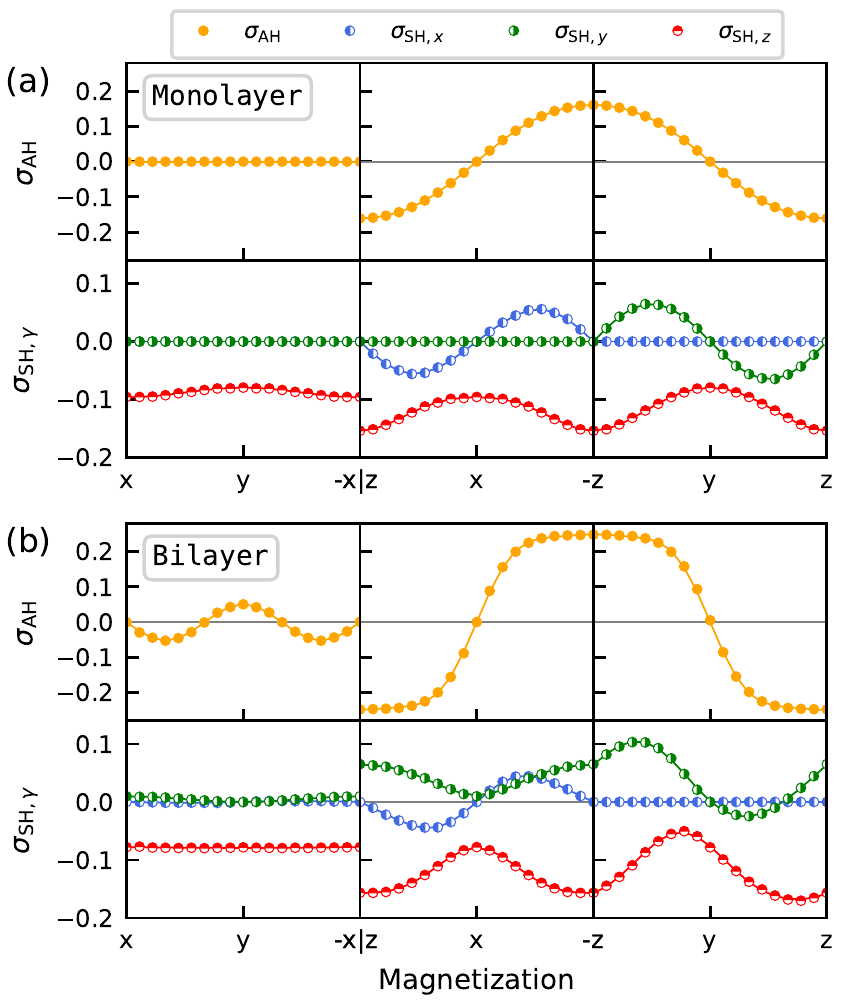}
	\caption{\label{fig:two}
		AHC and SHC of
		(a) monolayer and
		(b) bilayer FGT with magnetization
		rotating inside \textit{xy}-, \textit{xz}-, and \textit{yz}-planes.
		$\sigma_{\text{AH}}  $ of \textit{xy}-magnetization is magnified 10 times in (b).
		Markers denote the \data, while lines  depict  the fitting curves.
		SHC has been multiplied by
		a factor of $ -2e/\hbar $,
		thus the  units of  both
		$\sigma_{\text{AH}}  $ and
		$ \sigma_{\text{SH},\gamma} $
		are  \si{e^2/\hbar}.
	}
\end{figure}

The correlation between AHC and \textit{intra} SHC is discussed as follows.
It has been reported that AHC is  dominated by $\Omega(\bm{k})$ spikes
at  limited  $\bm{k}$-points,
where the spin-orbit-split bands
cross the  Fermi energy  $ E_F $ \cite{PhysRevLett.92.037204,PhysRevB.74.195118}.
These spikes  are caused by
the small energy gaps
between the \textit{n}th and the \textit{m}th bands
(one occupied and one unoccupied),
which lead  to  small energy denominators
	[$ (\epsilon_{n\bm{k}}-\epsilon_{m\bm{k}})^2 $].
Similar to AHC,
the dominant contribution to  {\textit{intra} SHC 
also  comes from  spikes in BZ.
Moreover,  
at these $\bm{k}$-points, 
bands have analogous spin \cite{supp},
indicating that   around  these $ \bar{\Omega}_{\gamma} (\bm{k}) $ spikes,
$  \langle n \bm{k}| \hat{\sigma}_\gamma  | n \bm{k}   \rangle \approx
	\langle m \bm{k}| \hat{\sigma}_\gamma  | m \bm{k}   \rangle $,
making  Eq.~(\ref{eq:spinberry-band})   approximated as

\begin{align}
	\begin{split}
		\label{eq:spinberry-band-appr}
		\bar \Omega_{n, \gamma}(\bm{k})& \approx
		\langle  n\bm{k}| \hat{\sigma}_\gamma    |  n\bm{k}   \rangle	\\
		&  \times {\hbar}^2 \sum_{	m\ne n}
		\frac{-2\operatorname{Im}[
				\langle  n\bm{k}| 	\hat{v}_x    	|m\bm{k}\rangle
				\langle m\bm{k}| \hat{v}_{y}|n\bm{k}\rangle]}
		{(\epsilon_{n\bm{k}}-\epsilon_{m\bm{k}})^2} \\
		& =  \langle  n\bm{k}| \hat{\sigma}_\gamma    |  n\bm{k}   \rangle
		\Omega_{n}(\bm{k}).\\
	\end{split}
\end{align}

\textcolor{black}
{
	Equation~(\ref{eq:spinberry-band-appr}) illustrates  that
	the spin of eigenstate
	is the bridge to connect
	Berry curvature
	and  \textit{intra} spin Berry curvature.
}
On the other hand,
only considering the direction,
$ \mathbf{\hat{m}} = \frac{ \textbf{m}}{ |\textbf{m}|}  $  is used to denote the unit vector of magnetization.
This unit vector is  projected onto  three axes
as  $ \mathbf{\hat{m}} =[m_x, m_y,m_z] $,
and
\begin{align}
	\label{eq:magnetization}
	m_{\gamma}  \sim
	\int_\text{BZ}  \frac{d^2 {k}}{(2\pi)^2}  \sum_{n}   f_{n\bm{k}}
	\langle n\bm{k}| \hat{\sigma}_\gamma  | n \bm{k}   \rangle .
\end{align}
Considering  Eqs. (\ref{eq:berry-band}),  (\ref{eq:spinberry-band-appr}), and (\ref{eq:magnetization}),
AHC and  \textit{intra} SHC  
can be related  by the following formula
\begin{equation}
	\label{eq:sahc}
	\bm \sigma_{\text{SH}}^{intra}  \approx \eta \sigma_{\text{AH}} \mathbf{\hat{m}},
\end{equation}
where
$ \bm \sigma_{\text{SH}}^{intra}  = [ \sigma_{\text{SH}, x}^{intra}  ,  \sigma_{\text{SH}, y}^{intra}, \sigma_{\text{SH}, z}^{intra} ] $,
the coefficient  $ \eta  $  denotes the conversion efficiency
from   AHC  to \textit{intra} SHC.
	\textcolor{black}{It should be mentioned that 
	$ \bm \sigma_{\text{SH}}^{intra}$, $ \eta  $, $ \sigma_{\text{AH}} $
	depend on $ \mathbf{\hat{m}} $, which is omitted in Eq.~(\ref{eq:sahc}) for brevity.}
The  approximation
is due to  that  $ \langle n\bm{k}| \hat{\sigma}_\gamma  | n \bm{k}   \rangle $
is not a good quantum number,
and it will be shown that  this approximation is negligible
in the case of Eq.~(\ref{eq:sahc}).
\textcolor{black}{
	More significantly,  
	$ \bm \sigma_{\text{SH}}^{intra} $ can be regarded as
	the intrinsic SAHE,
	since it denotes the spin current converted from AHE,
	and its  spin orientation  is along the magnetization.
	In a nutshell,
	the rigorous deviations of SAHE are presented using Kubo formula [Eqs.~(\ref{eq:spinberry-band})-(\ref{eq:kubo_sahc})],
	making  AHE and SAHE correlated   through the  magnetization
	within the approximation $  \langle n \bm{k}| \hat{\sigma}_\gamma  | n \bm{k}   \rangle \approx \langle m \bm{k}| \hat{\sigma}_\gamma  | m \bm{k}   \rangle $.
}

In the above section,
the diagonal elements of spin matrix has been considered  in order to define   \textit{intra} SHC.
Meanwhile,  the off-diagonal part  also contributes  to spin current.
Through  inserting
$\sum_{t}|t\bm{k}\rangle{\langle}t\bm{k}|=1$
into Eq.~(\ref{eq:berrylike-band})
and
considering 
$ t\ne n, m,$ respectively,
the off-diagonal  part,
called  \textcolor{black}{\textit{inter} spin Berry curvature},
is defined as
\begin{align}
	\begin{split}
		\label{eq:nondiag}
		& \tilde \Omega_{n, \gamma}(\bm{k}) =  \\
		&
		{\hbar}^2 \sum_{\substack{m\ne n\\t\ne m,n}}  -\operatorname{Im}
		\Big[ \langle n\bm{k}| \hat{\sigma}_\gamma| t\bm{k} \rangle \times
			\frac{
				\langle  t\bm{k}| 	\hat{v}_x |m\bm{k}\rangle
				\langle m\bm{k}| \hat{v}_{y}|n\bm{k}\rangle }
			{ (\epsilon_{n\bm{k}}-\epsilon_{m\bm{k}})^2} \\
			&  \hspace{1.8cm}+
			\langle t \bm{k}| \hat{\sigma}_\gamma| m\bm{k} \rangle \times
			\frac{
				\langle  n\bm{k}| 	\hat{v}_x |t\bm{k}\rangle
				\langle m\bm{k}| \hat{v}_{y}|n\bm{k}\rangle }
			{ (\epsilon_{n\bm{k}}-\epsilon_{m\bm{k}})^2}  \Big].\\
	\end{split}
\end{align}

\textcolor{black}{
	With this quantity and its summation,
	\textit{inter} spin  Hall conductivity 
	is  defined  as  
	\begin{equation}
		\tilde{\Omega}_{\gamma} (\bm{k})
		= \sum_{n}  f_{n\bm{k}}  \tilde{\Omega}_{n, \gamma}(\bm{k}),
	\end{equation}
	\begin{equation}
		\sigma_{\text{SH},\gamma}^{inter}   =
		-\frac{e^2}{\hbar}
		\int_\text{BZ}
		\frac{d^2 {k}}{(2\pi)^2}
		\tilde\Omega_{\gamma}  (\bm{k}).
	\end{equation} 
For the sake of intelligibility,
all of the terminologies and symbols 
used above
 are summarized
in Supplemental Material \cite{supp}.
}

\textit{AHC and SHC.}---The
\textit{ab initio} calculations have been performed to obtain AHC and SHC with
rotating magnetization as depicted in Fig.~\ref{fig:structure}(d),
and the results at  $ E_F $  are  summarized in Fig.~\ref{fig:two}.
Although
it is conventional to set
$ \mathbf{\hat{m}}  $  along \textit{z}-axis to maximize AHC,
indeed,
AHC is dependent on   $ \mathbf{\hat{m}}  $ \cite{PhysRevLett.103.097203}.
Breaking the
time-reversal symmetry
($ \mathcal{T} $)  \cite{1322762},
AHC shall be  expressed by
the  odd-order terms  with respect to magnetization.
On the contrary,
SHC
is invariant under  $ \mathcal{T} $  \cite{1322762}
and described by
even-order terms.
The fitting curves of the \data~are obtained  using   SciPy package \cite{scipy},
and  the complete expressions can be found in Supplemental Material \cite{supp}.

Figure~\ref{fig:two}(a) presents  the AHC and SHC  of monolayer FGT.
It can be observed that $  \sigma_\text{AH} = 0  $
when the  magnetization is in the \xyp.
In contrast,
both the \textit{xz}- and \textit{yz}-magnetization  cases
allow   the occurrence of AHC.
When $ \mathbf{\hat{m}}  $  is along $-z$,
the magnitude of AHC reaches  the maximum, 0.134 \si{e^2/\hbar},
larger than other vdW ferromagnet  \cite{PhysRevB.103.024436}.
According to the   directions of  spin polarization,
$ \bm\sigma_\text{SH}  $   is  decomposed into
$ \sigma_{\text{SH},x} , \sigma_{\text{SH},y}, \sigma_{\text{SH},z}  $.
Compared with the nonmagnetic transition metal dichalcogenides  \cite{PhysRevB.86.165108},
the \ml FGT can exhibit much larger
SHC.
Besides,
the SHC in FGT is nonlinear and periodic with respect to $ \mathbf{\hat{m}} $,
indicating  the magnitude of spin current  can be controlled by magnetization.
More intriguingly,
various tensors emerge in FGT,
making  this system  more practical for  applications.
For instance, when the magnetization is in the \xzp,
$ \sigma_{\text{SH},x} $ and
$ \sigma_{\text{SH},z} $  tensors
appear simultaneously,
making  the spin current
induce both  in-plane and out-of-plane torques.

Different from the monolayer,
the \bl FGT,
with     $ \mathcal{P} $ symmetry,
can  exhibit  non-zero $ \sigma_\text{AH} $  with a period of $ 2\pi/3 $
in the case of \textit{xy}-\magn~
[see  Fig.~\ref{fig:two}(b)].
This can be explained by Neumann’s principle,
which states that
the symmetry of physical property must include all the symmetries of  crystal \cite{Newnham2005Jan, Gallegolk5043,PhysRevB.99.104428}.
Since  the magnetization is a pseudovector,
when $ \mathbf{\hat{m}}  $ is orthogonal  to any   mirror shown in 
Fig.~\ref{fig:structure}(c),
$ \mathcal{M} $  would be preserved, thus prohibiting
the occurrence of  AHC.
Consequently,
$ \sigma_\text{AH} = 0 $ occurs  periodically.
When   $ \mathbf{\hat{m}} $ points to other orientations,
mirrors would be broken
and not  restrict  AHC.
Besides, compared with the monolayer,
a larger  AHC of  0.248 \si{e^2/\hbar} is found in the \bl FGT.
It is worth  noting that more diverse SHC tensors
are  found in the \bl FGT.
In the \textit{xz}-\magn~scenario,
$ \sigma_{\text{SH},x} $,  $ \sigma_{\text{SH},y} $, and $ \sigma_{\text{SH},z} $
tensors  can all occur simultaneously.
The existence of   $ \sigma_{\text{SH},y} $  in \bl FGT
is    distinct  from that in  \ml FGT.
The reason is attributed to the  difference in symmetry,
i.e.,
the \textit{xz}-magnetization  \ml   preserves
the  $ \mathcal{TC} $
($ \mathcal{C} $ denotes  the rotational symmetry)
while the \bl not,
thus the latter can stimulate  peculiar SHC tensor.
The present findings  evidently  demonstrate  that
the interplay between  magnetism  and  symmetry
is effective  to manipulate  the
magnitude and polarization  of
spin current.

\begin{figure}[tbh]
	\includegraphics[scale=1]{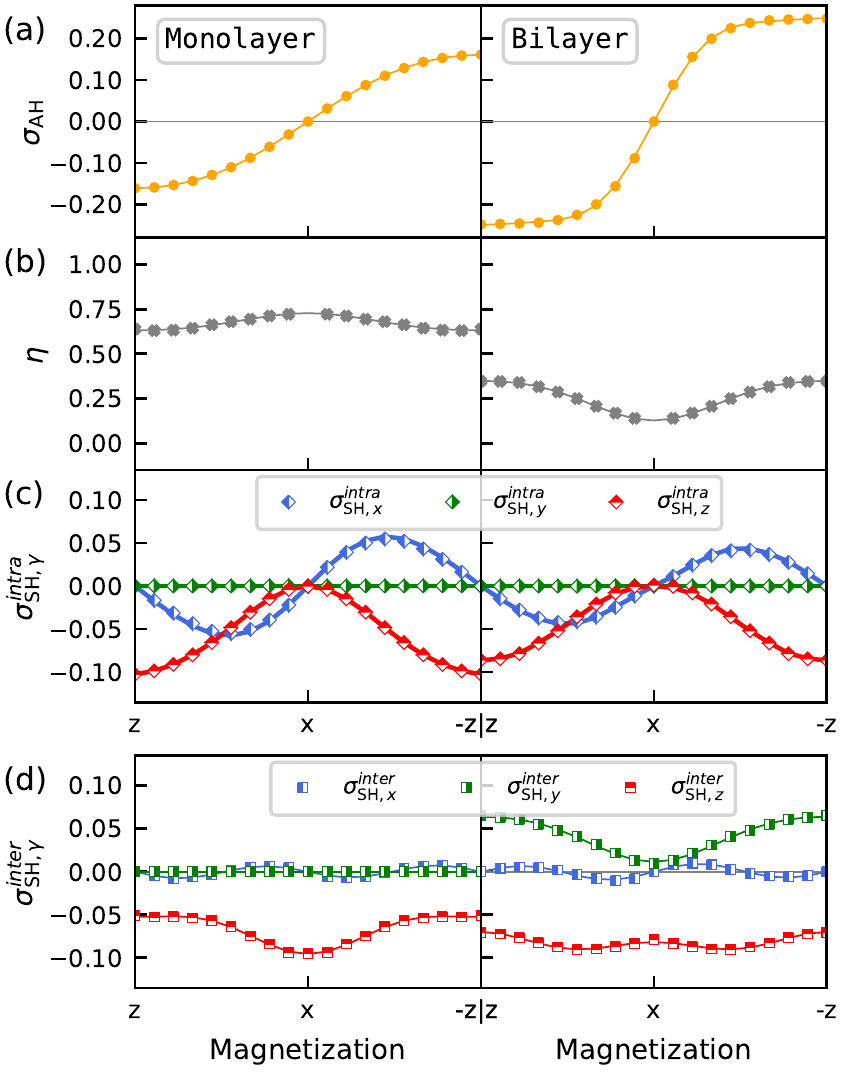}
	\caption{\label{fig:sahc}
		(a) AHC,
		(b) conversion efficiency,
		(c) \textit{intra} SHC,
		and  (d) \textit{inter} SHC
		of \ml and \bl  FGT, respectively,
		with the \textit{xz}-magnetization.
		The colorful markers in (a), (c), and (d) denote the \data, 
		while the grey markers in (b) are derived data.
		(a), (b), and (d) depict  the fitting curves (slim lines), 
		while (c) depicts the curves of   Eq.~(\ref{eq:sahc})
		(bold lines).
		The units of  $\sigma_{\text{AH}}$,
		$ \sigma_{\text{SH},\gamma}^{intra} $, and
		$ \sigma_{\text{SH},\gamma}^{inter} $  are  \si{e^2/\hbar}.
	}
\end{figure}

\begin{figure*}[htb]
	\includegraphics[scale=0.67]{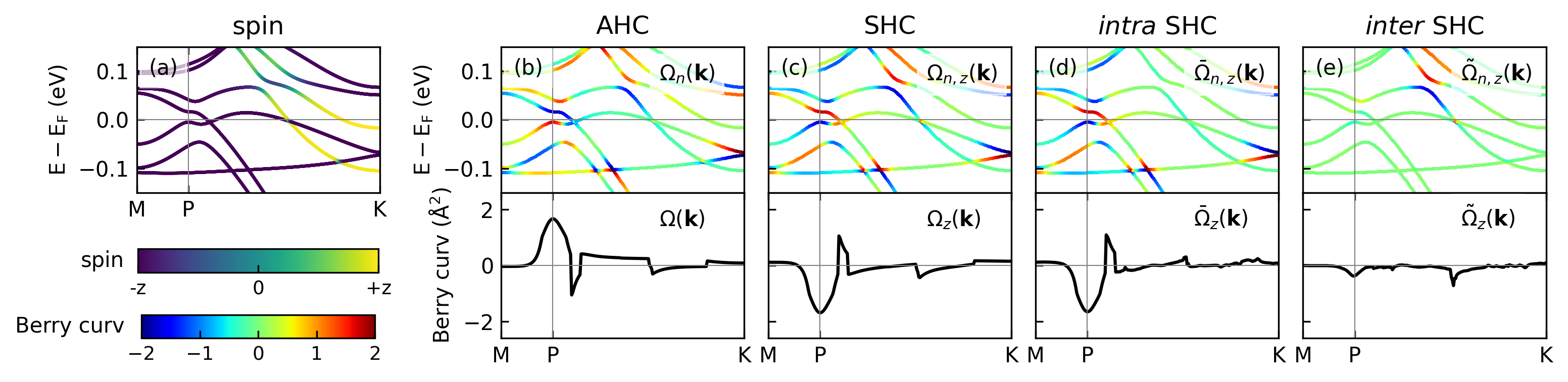}
	\caption{\label{fig:kpath} \textcolor{black}{
(a) Spin projected bands of bilayer FGT with \textit{z}-magnetization,
the dark (bright) color denotes spin along $ -z $ ($ +z $). 
(b) Berry curvature, 
(c) spin Berry curvature,
(d) \textit{intra} spin Berry curvature,
and
(e) \textit{inter} spin Berry curvature.
The upper panels show the Berry curvatures projected bands,
and 
the lower panels depict the  summation results.
All the Berry curvatures are in log scale [Eq. S9],
the blue (red) color denotes the negative (positive) curvature value.
The titles indicate the corresponding conductivities.  
$ E_F $ is set to zero.}
	}
\end{figure*}

\textcolor{black}{\textit{Conversion from  AHC to  SAHC.}---It
	has been discussed that
	SHC  can be decomposed into  {\textit{intra} SHC} and  {\textit{inter} SHC},
	and the {\textit{intra} SHC} (i.e. SAHC) depends on AHC.
	The conversion from  AHC to  \textit{intra} SHC
	is essential to comprehend \textit{intra} SHC.}
Through \fp,  the AHC and  \textit{intra} SHC
with \textit{xz}-magnetization have been investigated
	[see markers in Figs.~\ref{fig:sahc}(a) and (c)],
note the \textit{intra} SHC was  calculated
using the \textit{intra} spin Berry curvature
	[Eqs.~(\ref{eq:spinberry-band})-(\ref{eq:kubo_sahc})].
Using Eq.~\ref{eq:sahc}, the conversion efficiency $ \eta $ from  AHC  to  \textit{intra} SHC
can be  derived by
$ \sigma_{\text{SH},\gamma}^{intra}  / (m_\gamma \sigma_\text{AH}) $,
and 
$ \eta $ is
displayed by markers in Fig.~\ref{fig:sahc}(b).
Using  fitting functions of  AHC and $ \eta  $, 
the evolution   
of \textit{intra} SHC with respect to $\mathbf{\hat{m}}  $ 
can be obtained
using Eq.~(\ref{eq:sahc}).
Figure~\ref{fig:sahc}(c) illustrates that in both \ml and \bl FGT,
the evolution curves  of  Eq.~(\ref{eq:sahc}) [see bold lines]  can
perfectly  characterize
the \textit{ab initio}  data  of  \textit{intra} SHC 
for any tensor element
and any magnetization direction,
demonstrating that 
the set of conversion efficiency is universal.
Moreover, the case  of
\textit{yz}-magnetization
draws  the same conclusion 
with the same $ \eta $ parameters \cite{supp}.
Thus,
the correlation between AHC and \textit{intra} SHC,
i.e.   Eq.~(\ref{eq:sahc}), is  verified.
Indeed,
the anomalous Hall current is always  accompanied by
spin anomalous Hall current,
with a spin orientation collinear with $\mathbf{\hat{m}} $ \cite{PhysRevApplied.3.044001}.
Consequently,
\textit{intra} SHC is expected to be dependent on AHC.
The  multiplier   $ \mathbf{\hat{m}} =[m_x, m_y,m_z] $  can be
regarded as  the  projection
onto \textit{x}-, \textit{y}-, and \textit{z}-direction,
producing three  spin current components
$\sigma_{\text{SH}, x}^{intra}, \sigma_{\text{SH}, y}^{intra},\sigma_{\text{SH}, z}^{intra} $,
respectively.

\textcolor{black}{Different from the  \textit{intra} SHC,
	there is no universal conversion efficiency
	from AHC to \textit{inter} SHC, which is shown in the separated
	Fig.~\ref{fig:sahc}(d).}
It is worth  mentioning  that
even if the spin  orientates inside the  \xzp,
\textit{inter}  SHC can still contribute to the \textit{y}-polarization
(non-zero tensor  $  \sigma_{\text{SH}, y}^{inter} $) in   \bl  FGT,
i.e., the \textit{y}-polarization SHC
is exclusively contributed by \textit{inter} SHC.

\textcolor{black}
{
	\textit{Berry curvatures.}---The 
	band structures of \bl FGT
	have been investigated to
	present the microscopic mechanism
	of the Hall conductivities.
	The  Te-\textit{p} and Fe-\textit{d}  orbitals make the dominant contributions to
	the bands around Fermi energy \cite{supp}.
	The projected bands along M-K
	are selected  to clarify
	the correlations  of  various Berry curvatures.
	Figure~\ref{fig:kpath}(a)  shows that at $ E_F $, the  spin orientations
	are well (anti-)aligned  with the magnetization.
	Figures~\ref{fig:kpath}(b)-(e) present
	the \textit{ab initio} results of various Berry curvatures.
	It can be seen that 
	main contributions to Berry curvatures 
	come from the band pairs
	with  small energy gaps.
	As expected, for both  $  \Omega(\bm{k} )  $  and  $ \bar\Omega_{z}(\bm{k})$,
	large  spikes only  occur
	at  limited  $\bm{k}$-points such as the P point,
	where the bands have analogous spin.
	On the contrary,
	$ \tilde\Omega_{z}(\bm{k})$ is mainly contributed by the bands
	with misaligned spin away from $ E_F $.
}

\textcolor{black}
{
	The conversion from AHC to \textit{intra} SHC
	has been studied from  the  conductivity aspect 
	in Fig. \ref{fig:sahc}.
	To reveal the microscopic mechanisms,
	the conversion
	from Berry curvature
	to  \textit{intra} spin Berry curvature
	would be discussed  at two levels:
	(i) The band-resolved conversion efficiency
	$ \eta_n(\bm{k})=\bar\Omega_{n,z}(\bm{k}) / \Omega_n (\bm{k})$,
	and
	(ii) the summation result
	$ \eta(\bm{k}) =  \bar\Omega_{z}(\bm{k}) / \Omega(\bm{k})  $.
	Figures~\ref{fig:kpath}(a), (b), and (d) demonstrate that
	the  $\eta_n(\bm{k})$
	is determined by the spin of eigenstate,  making
	$\eta_n(\bm{k}) = \langle n\bm{k}|\hat{\sigma}_z|n\bm{k} \rangle = \pm 1$  
	when  the spin is collinear  with the magnetization.
	For instance,
	two bands around $ E_F $ (one occupied and one unoccupied) at P point
	have spin along $ -z $,
	i.e. $\eta_n(\bm{k}) =-1$, leading to the opposite signs of $ \Omega_n(\bm{k})$ and $\bar\Omega_{n,z}(\bm{k})$ on these two bands,
	as shown in the upper panels of
	Figs.~\ref{fig:kpath}(b) and (d).
	On the other hand,
	although  $\Omega(\bm{k})$   [$\bar\Omega_z(\bm{k})$]
	is a property of the occupied manifold,
	it is mostly determined by the band structure nearby
	Fermi surface,
	since the contributions of bands far away  from  $ E_F $
	are cancelled by each other  \cite{PhysRevB.76.195109, PhysRevLett.100.096401}.
	Consequently, $ \eta(\bm{k}) $, to some extent,
	is  determined by the spin of occupied state around $ E_F $.
	For instance, at P point, the last occupied band exhibits spin along $ -z $,
	resulting in  $ \eta(\bm{k})  = -1$
	as shown in the lower panels of
	Figs.~\ref{fig:kpath}(b) and (d).
	Note that  the conductivities 
	$ \sigma_{\text{AH}} $ and $\sigma_{\text{SH},z}^{intra} $ are,
	respectively,
	the integral of $\Omega(\bm{k})$ and $ \bar\Omega_{z}(\bm{k})$
	with respect to $ \bm{k} $.
	However,
	it should be emphasized that
	$ \eta $ is not the integral of $\eta(\bm{k})$ with respect to $ \bm{k} $,
	since neither $\Omega(\bm{k})$, $\bar\Omega_{z}(\bm{k})$
	nor $\eta(\bm{k})$ is a constant
	in the Brillouin zone of FGT \cite{supp}.
	In fact,
	there is no simple relation
	between the global conversion efficiency    $ \eta $
	and the microscopic $ \eta_n(\bm{k})  $
	in the  FGT system.
	Thus,
	the global  $ \eta $
	shall be obtained using  the global conductivities 
	AHC and \textit{intra} SHC,
	and the value of  $ \eta $  is not restricted
	in the range of  $ [-1, 1] $.
}
 
\textcolor{black}
{
	\textit{Effective model.}---An effective
	$ k\cdot p $  model
	is constructed to reveal more physical principles.
	Under the  basis  of
	$[c_{1,k\uparrow}, c_{2,k\uparrow}, c_{1,k\downarrow}, c_{2,k\downarrow}]^T $
	where 1 and 2 denote orbital index, the $ 4\times4 $ Hamiltonian can
	be expressed as
	\begin{align}
		\begin{split}
			H ={ }   &   (mk^2 + \delta)\tau_z \\
			& + \alpha k_x \sigma_z \otimes \tau_x
			+ \beta  k_y \sigma_0 \otimes \tau_y
			+ M \sigma_z \otimes \tau_0,  \\
		\end{split}
	\end{align}
	where
	$ \sigma_i $  and  $ \tau_i $
	are Pauli matrices
	for the spin and orbital degrees of freedom, respectively.
	$ m $ denotes the effective mass,
	$ \delta $ is the strength of band inversion.
	$ \alpha $ and $ \beta $ terms respectively 
	denote   the SOC and the orbital  hybridization.
	The last term indicates a magnetic field along $ z $ direction.
	This model   qualitatively
	characterizes the \textit{ab initio} bands around  P point  and  $ E_F $
	shown in Fig. \ref{fig:kpath},
	as well as  the properties of spin and Berry curvatures.
	Moreover,
	it   reveals that
	both $ \Omega_n $  and  $ \bar\Omega_{n,z}$ are
	proportional to the strength of the orbital  hybridization
	while
	inversely  proportional to SOC and band inversion \cite{supp}.
}

\textit{Summary.}---The origin and   control  of  spin current  in  both  \ml and \bl    Fe$_3$GeTe$_2$
have  been systematically investigated.
It exhibits nonlinear  behavior  with respect to magnetization,
as well as the simultaneous occurrence of
in-plane and out-of-plane spin polarizations.
Superior to the  monolayer case,
\bl  Fe$_3$GeTe$_2$
can present  unusual arbitrary spin current tensor
due to the reduced symmetry.
Using the concepts  of  Berry   curvatures
and \textit{intra} spin Berry  curvature,
the correlation
between anomalous Hall effect
and spin anomalous Hall effect
has been clarified,
as well as  the corresponding  conversion efficiency.
\textcolor{black}{An effective $ k \cdot p $ model
illustrates that the orbital hybridization
is essential for Berry curvatures.}
The present research  demonstrates that
the interplay between  magnetism  and  symmetry
can  control  both   magnitude and polarization of the spin current,
further  stimulating  exotic  spin-orbit torques for spintronic devices.

\textit{Acknowledgments.}---Jiaqi Zhou acknowledges fruitful discussions with Junfeng Qiao. The authors acknowledge financial support from the F\'{e}d\'{e}ration Wallonie-Bruxelles through the ARC entitled ``3D nanoarchitecturing of 2D crystals" project (ARC 16/21-077), from the European Union's Horizon 2020 research and innovation programme (Core3 - N° 881603), from the Flag-ERA JTC 2019 project entitled ``SOGraphMEM" (R.8012.19),
and from the Belgium F.R.S-FNRS under the conventions N° T.0051.18. Computational resources have been provided by the supercomputing facilities of the Universit\'{e} catholique de Louvain (CISM) and the Consortium des Equipements de Calcul Intensif en F\'{e}d\'{e}ration Wallonie Bruxelles (CECI) funded by the Fonds de la Recherche Scientifique de Belgique (F.R.S-FNRS - N° 2.5020.11).


\begin{thebibliography}{49}%
	\makeatletter
	\providecommand \@ifxundefined [1]{%
		\@ifx{#1\undefined}
	}%
	\providecommand \@ifnum [1]{%
		\ifnum #1\expandafter \@firstoftwo
		\else \expandafter \@secondoftwo
		\fi
	}%
	\providecommand \@ifx [1]{%
		\ifx #1\expandafter \@firstoftwo
		\else \expandafter \@secondoftwo
		\fi
	}%
	\providecommand \natexlab [1]{#1}%
	\providecommand \enquote  [1]{``#1''}%
	\providecommand \bibnamefont  [1]{#1}%
	\providecommand \bibfnamefont [1]{#1}%
	\providecommand \citenamefont [1]{#1}%
	\providecommand \href@noop [0]{\@secondoftwo}%
	\providecommand \href [0]{\begingroup \@sanitize@url \@href}%
	\providecommand \@href[1]{\@@startlink{#1}\@@href}%
	\providecommand \@@href[1]{\endgroup#1\@@endlink}%
	\providecommand \@sanitize@url [0]{\catcode `\\12\catcode `\$12\catcode
		`\&12\catcode `\#12\catcode `\^12\catcode `\_12\catcode `\%12\relax}%
	\providecommand \@@startlink[1]{}%
	\providecommand \@@endlink[0]{}%
	\providecommand \url  [0]{\begingroup\@sanitize@url \@url }%
	\providecommand \@url [1]{\endgroup\@href {#1}{\urlprefix }}%
	\providecommand \urlprefix  [0]{URL }%
	\providecommand \Eprint [0]{\href }%
	\providecommand \doibase [0]{https://doi.org/}%
	\providecommand \selectlanguage [0]{\@gobble}%
	\providecommand \bibinfo  [0]{\@secondoftwo}%
	\providecommand \bibfield  [0]{\@secondoftwo}%
	\providecommand \translation [1]{[#1]}%
	\providecommand \BibitemOpen [0]{}%
	\providecommand \bibitemStop [0]{}%
	\providecommand \bibitemNoStop [0]{.\EOS\space}%
	\providecommand \EOS [0]{\spacefactor3000\relax}%
	\providecommand \BibitemShut  [1]{\csname bibitem#1\endcsname}%
	\let\auto@bib@innerbib\@empty
	\bibitem [{\citenamefont {Dieny}\ \emph {et~al.}(2020)\citenamefont {Dieny},
		\citenamefont {Prejbeanu}, \citenamefont {Garello}, \citenamefont
		{Gambardella}, \citenamefont {Freitas}, \citenamefont {Lehndorff},
		\citenamefont {Raberg}, \citenamefont {Ebels}, \citenamefont {Demokritov},
		\citenamefont {Akerman}, \citenamefont {Deac}, \citenamefont {Pirro},
		\citenamefont {Adelmann}, \citenamefont {Anane}, \citenamefont {Chumak},
		\citenamefont {Hirohata}, \citenamefont {Mangin}, \citenamefont {Valenzuela},
		\citenamefont
		{Onba{\ifmmode\mbox{\c{s}}\else\c{s}\fi}l{\ifmmode\imath\else\i\fi}},
		\citenamefont {D{'}Aquino}, \citenamefont {Prenat}, \citenamefont
		{Finocchio}, \citenamefont {Lopez-Diaz}, \citenamefont {Chantrell},
		\citenamefont {Chubykalo-Fesenko},\ and\ \citenamefont
		{Bortolotti}}]{Dieny2020Aug}%
	\BibitemOpen
	\bibfield  {author} {\bibinfo {author} {\bibfnamefont {B.}~\bibnamefont
			{Dieny}}, \bibinfo {author} {\bibfnamefont {I.~L.}\ \bibnamefont
			{Prejbeanu}}, \bibinfo {author} {\bibfnamefont {K.}~\bibnamefont {Garello}},
		\bibinfo {author} {\bibfnamefont {P.}~\bibnamefont {Gambardella}}, \bibinfo
		{author} {\bibfnamefont {P.}~\bibnamefont {Freitas}}, \bibinfo {author}
		{\bibfnamefont {R.}~\bibnamefont {Lehndorff}}, \bibinfo {author}
		{\bibfnamefont {W.}~\bibnamefont {Raberg}}, \bibinfo {author} {\bibfnamefont
			{U.}~\bibnamefont {Ebels}}, \bibinfo {author} {\bibfnamefont {S.~O.}\
			\bibnamefont {Demokritov}}, \bibinfo {author} {\bibfnamefont
			{J.}~\bibnamefont {Akerman}}, \bibinfo {author} {\bibfnamefont
			{A.}~\bibnamefont {Deac}}, \bibinfo {author} {\bibfnamefont {P.}~\bibnamefont
			{Pirro}}, \bibinfo {author} {\bibfnamefont {C.}~\bibnamefont {Adelmann}},
		\bibinfo {author} {\bibfnamefont {A.}~\bibnamefont {Anane}}, \bibinfo
		{author} {\bibfnamefont {A.~V.}\ \bibnamefont {Chumak}}, \bibinfo {author}
		{\bibfnamefont {A.}~\bibnamefont {Hirohata}}, \bibinfo {author}
		{\bibfnamefont {S.}~\bibnamefont {Mangin}}, \bibinfo {author} {\bibfnamefont
			{S.~O.}\ \bibnamefont {Valenzuela}}, \bibinfo {author} {\bibfnamefont
			{M.~C.}\ \bibnamefont
			{Onba{\ifmmode\mbox{\c{s}}\else\c{s}\fi}l{\ifmmode\imath\else\i\fi}}},
		\bibinfo {author} {\bibfnamefont {M.}~\bibnamefont {D{'}Aquino}}, \bibinfo
		{author} {\bibfnamefont {G.}~\bibnamefont {Prenat}}, \bibinfo {author}
		{\bibfnamefont {G.}~\bibnamefont {Finocchio}}, \bibinfo {author}
		{\bibfnamefont {L.}~\bibnamefont {Lopez-Diaz}}, \bibinfo {author}
		{\bibfnamefont {R.}~\bibnamefont {Chantrell}}, \bibinfo {author}
		{\bibfnamefont {O.}~\bibnamefont {Chubykalo-Fesenko}},\ and\ \bibinfo
		{author} {\bibfnamefont {P.}~\bibnamefont {Bortolotti}},\ }\href
	{https://doi.org/10.1038/s41928-020-0461-5} {\bibfield  {journal} {\bibinfo
			{journal} {Nat. Electron.}\ }\textbf {\bibinfo {volume} {3}},\ \bibinfo
		{pages} {446} (\bibinfo {year} {2020})}\BibitemShut {NoStop}%
	\bibitem [{\citenamefont {Peng}\ \emph {et~al.}(2019)\citenamefont {Peng},
		\citenamefont {Zhu}, \citenamefont {Zhou}, \citenamefont {Zhang},
		\citenamefont {Cao}, \citenamefont {Wang}, \citenamefont {Cai}, \citenamefont
		{Cao},\ and\ \citenamefont {Zhao}}]{aelm.201900134}%
	\BibitemOpen
	\bibfield  {author} {\bibinfo {author} {\bibfnamefont {S.}~\bibnamefont
			{Peng}}, \bibinfo {author} {\bibfnamefont {D.}~\bibnamefont {Zhu}}, \bibinfo
		{author} {\bibfnamefont {J.}~\bibnamefont {Zhou}}, \bibinfo {author}
		{\bibfnamefont {B.}~\bibnamefont {Zhang}}, \bibinfo {author} {\bibfnamefont
			{A.}~\bibnamefont {Cao}}, \bibinfo {author} {\bibfnamefont {M.}~\bibnamefont
			{Wang}}, \bibinfo {author} {\bibfnamefont {W.}~\bibnamefont {Cai}}, \bibinfo
		{author} {\bibfnamefont {K.}~\bibnamefont {Cao}},\ and\ \bibinfo {author}
		{\bibfnamefont {W.}~\bibnamefont {Zhao}},\ }\href
	{https://doi.org/https://doi.org/10.1002/aelm.201900134} {\bibfield
		{journal} {\bibinfo  {journal} {Adv. Electron. Mater.}\ }\textbf {\bibinfo
			{volume} {5}},\ \bibinfo {pages} {1900134} (\bibinfo {year}
		{2019})}\BibitemShut {NoStop}%
	\bibitem [{\citenamefont {Ikeda}\ \emph {et~al.}(2010)\citenamefont {Ikeda},
		\citenamefont {Miura}, \citenamefont {Yamamoto}, \citenamefont {Mizunuma},
		\citenamefont {Gan}, \citenamefont {Endo}, \citenamefont {Kanai},
		\citenamefont {Hayakawa}, \citenamefont {Matsukura},\ and\ \citenamefont
		{Ohno}}]{Ikeda2010Sep}%
	\BibitemOpen
	\bibfield  {author} {\bibinfo {author} {\bibfnamefont {S.}~\bibnamefont
			{Ikeda}}, \bibinfo {author} {\bibfnamefont {K.}~\bibnamefont {Miura}},
		\bibinfo {author} {\bibfnamefont {H.}~\bibnamefont {Yamamoto}}, \bibinfo
		{author} {\bibfnamefont {K.}~\bibnamefont {Mizunuma}}, \bibinfo {author}
		{\bibfnamefont {H.~D.}\ \bibnamefont {Gan}}, \bibinfo {author} {\bibfnamefont
			{M.}~\bibnamefont {Endo}}, \bibinfo {author} {\bibfnamefont {S.}~\bibnamefont
			{Kanai}}, \bibinfo {author} {\bibfnamefont {J.}~\bibnamefont {Hayakawa}},
		\bibinfo {author} {\bibfnamefont {F.}~\bibnamefont {Matsukura}},\ and\
		\bibinfo {author} {\bibfnamefont {H.}~\bibnamefont {Ohno}},\ }\href
	{https://doi.org/10.1038/nmat2804} {\bibfield  {journal} {\bibinfo  {journal}
			{Nat. Mater.}\ }\textbf {\bibinfo {volume} {9}},\ \bibinfo {pages} {721}
		(\bibinfo {year} {2010})}\BibitemShut {NoStop}%
	\bibitem [{\citenamefont {Ramaswamy}\ \emph {et~al.}(2018)\citenamefont
		{Ramaswamy}, \citenamefont {Lee}, \citenamefont {Cai},\ and\ \citenamefont
		{Yang}}]{Ramaswamy2018Sep}%
	\BibitemOpen
	\bibfield  {author} {\bibinfo {author} {\bibfnamefont {R.}~\bibnamefont
			{Ramaswamy}}, \bibinfo {author} {\bibfnamefont {J.~M.}\ \bibnamefont {Lee}},
		\bibinfo {author} {\bibfnamefont {K.}~\bibnamefont {Cai}},\ and\ \bibinfo
		{author} {\bibfnamefont {H.}~\bibnamefont {Yang}},\ }\href
	{https://doi.org/10.1063/1.5041793} {\bibfield  {journal} {\bibinfo
			{journal} {Appl. Phys. Rev.}\ }\textbf {\bibinfo {volume} {5}},\ \bibinfo
		{pages} {031107} (\bibinfo {year} {2018})}\BibitemShut {NoStop}%
	\bibitem [{\citenamefont {Manchon}\ \emph {et~al.}(2019)\citenamefont
		{Manchon}, \citenamefont {\ifmmode~\check{Z}\else \v{Z}\fi{}elezn\'y},
		\citenamefont {Miron}, \citenamefont {Jungwirth}, \citenamefont {Sinova},
		\citenamefont {Thiaville}, \citenamefont {Garello},\ and\ \citenamefont
		{Gambardella}}]{RevModPhys.91.035004}%
	\BibitemOpen
	\bibfield  {author} {\bibinfo {author} {\bibfnamefont {A.}~\bibnamefont
			{Manchon}}, \bibinfo {author} {\bibfnamefont {J.}~\bibnamefont
			{\ifmmode~\check{Z}\else \v{Z}\fi{}elezn\'y}}, \bibinfo {author}
		{\bibfnamefont {I.~M.}\ \bibnamefont {Miron}}, \bibinfo {author}
		{\bibfnamefont {T.}~\bibnamefont {Jungwirth}}, \bibinfo {author}
		{\bibfnamefont {J.}~\bibnamefont {Sinova}}, \bibinfo {author} {\bibfnamefont
			{A.}~\bibnamefont {Thiaville}}, \bibinfo {author} {\bibfnamefont
			{K.}~\bibnamefont {Garello}},\ and\ \bibinfo {author} {\bibfnamefont
			{P.}~\bibnamefont {Gambardella}},\ }\href
	{https://doi.org/10.1103/RevModPhys.91.035004} {\bibfield  {journal}
		{\bibinfo  {journal} {Rev. Mod. Phys.}\ }\textbf {\bibinfo {volume} {91}},\
		\bibinfo {pages} {035004} (\bibinfo {year} {2019})}\BibitemShut {NoStop}%
	\bibitem [{\citenamefont {Liu}\ \emph {et~al.}(2012)\citenamefont {Liu},
		\citenamefont {Lee}, \citenamefont {Gudmundsen}, \citenamefont {Ralph},\ and\
		\citenamefont {Buhrman}}]{PhysRevLett.109.096602}%
	\BibitemOpen
	\bibfield  {author} {\bibinfo {author} {\bibfnamefont {L.}~\bibnamefont
			{Liu}}, \bibinfo {author} {\bibfnamefont {O.~J.}\ \bibnamefont {Lee}},
		\bibinfo {author} {\bibfnamefont {T.~J.}\ \bibnamefont {Gudmundsen}},
		\bibinfo {author} {\bibfnamefont {D.~C.}\ \bibnamefont {Ralph}},\ and\
		\bibinfo {author} {\bibfnamefont {R.~A.}\ \bibnamefont {Buhrman}},\ }\href
	{https://doi.org/10.1103/PhysRevLett.109.096602} {\bibfield  {journal}
		{\bibinfo  {journal} {Phys. Rev. Lett.}\ }\textbf {\bibinfo {volume} {109}},\
		\bibinfo {pages} {096602} (\bibinfo {year} {2012})}\BibitemShut {NoStop}%
	\bibitem [{\citenamefont {Zhu}\ \emph {et~al.}(2019)\citenamefont {Zhu},
		\citenamefont {Ralph},\ and\ \citenamefont
		{Buhrman}}]{PhysRevLett.122.077201}%
	\BibitemOpen
	\bibfield  {author} {\bibinfo {author} {\bibfnamefont {L.}~\bibnamefont
			{Zhu}}, \bibinfo {author} {\bibfnamefont {D.~C.}\ \bibnamefont {Ralph}},\
		and\ \bibinfo {author} {\bibfnamefont {R.~A.}\ \bibnamefont {Buhrman}},\
	}\href {https://doi.org/10.1103/PhysRevLett.122.077201} {\bibfield  {journal}
		{\bibinfo  {journal} {Phys. Rev. Lett.}\ }\textbf {\bibinfo {volume} {122}},\
		\bibinfo {pages} {077201} (\bibinfo {year} {2019})}\BibitemShut {NoStop}%
	\bibitem [{\citenamefont {Sinova}\ \emph {et~al.}(2015)\citenamefont {Sinova},
		\citenamefont {Valenzuela}, \citenamefont {Wunderlich}, \citenamefont
		{Back},\ and\ \citenamefont {Jungwirth}}]{RevModPhys.87.1213}%
	\BibitemOpen
	\bibfield  {author} {\bibinfo {author} {\bibfnamefont {J.}~\bibnamefont
			{Sinova}}, \bibinfo {author} {\bibfnamefont {S.~O.}\ \bibnamefont
			{Valenzuela}}, \bibinfo {author} {\bibfnamefont {J.}~\bibnamefont
			{Wunderlich}}, \bibinfo {author} {\bibfnamefont {C.~H.}\ \bibnamefont
			{Back}},\ and\ \bibinfo {author} {\bibfnamefont {T.}~\bibnamefont
			{Jungwirth}},\ }\href {https://doi.org/10.1103/RevModPhys.87.1213} {\bibfield
		{journal} {\bibinfo  {journal} {Rev. Mod. Phys.}\ }\textbf {\bibinfo
			{volume} {87}},\ \bibinfo {pages} {1213} (\bibinfo {year}
		{2015})}\BibitemShut {NoStop}%
	\bibitem [{\citenamefont {Wang}\ \emph {et~al.}(2018)\citenamefont {Wang},
		\citenamefont {Cai}, \citenamefont {Zhu}, \citenamefont {Wang}, \citenamefont
		{Kan}, \citenamefont {Zhao}, \citenamefont {Cao}, \citenamefont {Wang},
		\citenamefont {Zhang}, \citenamefont {Zhang}, \citenamefont {Park},
		\citenamefont {Wang}, \citenamefont {Fert},\ and\ \citenamefont
		{Zhao}}]{Wang2018Nov}%
	\BibitemOpen
	\bibfield  {author} {\bibinfo {author} {\bibfnamefont {M.}~\bibnamefont
			{Wang}}, \bibinfo {author} {\bibfnamefont {W.}~\bibnamefont {Cai}}, \bibinfo
		{author} {\bibfnamefont {D.}~\bibnamefont {Zhu}}, \bibinfo {author}
		{\bibfnamefont {Z.}~\bibnamefont {Wang}}, \bibinfo {author} {\bibfnamefont
			{J.}~\bibnamefont {Kan}}, \bibinfo {author} {\bibfnamefont {Z.}~\bibnamefont
			{Zhao}}, \bibinfo {author} {\bibfnamefont {K.}~\bibnamefont {Cao}}, \bibinfo
		{author} {\bibfnamefont {Z.}~\bibnamefont {Wang}}, \bibinfo {author}
		{\bibfnamefont {Y.}~\bibnamefont {Zhang}}, \bibinfo {author} {\bibfnamefont
			{T.}~\bibnamefont {Zhang}}, \bibinfo {author} {\bibfnamefont
			{C.}~\bibnamefont {Park}}, \bibinfo {author} {\bibfnamefont {J.-P.}\
			\bibnamefont {Wang}}, \bibinfo {author} {\bibfnamefont {A.}~\bibnamefont
			{Fert}},\ and\ \bibinfo {author} {\bibfnamefont {W.}~\bibnamefont {Zhao}},\
	}\href {https://doi.org/10.1038/s41928-018-0160-7} {\bibfield  {journal}
		{\bibinfo  {journal} {Nat. Electron.}\ }\textbf {\bibinfo {volume} {1}},\
		\bibinfo {pages} {582} (\bibinfo {year} {2018})}\BibitemShut {NoStop}%
	\bibitem [{\citenamefont {Yu}\ \emph {et~al.}(2014)\citenamefont {Yu},
		\citenamefont {Upadhyaya}, \citenamefont {Fan}, \citenamefont {Alzate},
		\citenamefont {Jiang}, \citenamefont {Wong}, \citenamefont {Takei},
		\citenamefont {Bender}, \citenamefont {Chang}, \citenamefont {Jiang},
		\citenamefont {Lang}, \citenamefont {Tang}, \citenamefont {Wang},
		\citenamefont {Tserkovnyak}, \citenamefont {Amiri},\ and\ \citenamefont
		{Wang}}]{Yu2014Jul}%
	\BibitemOpen
	\bibfield  {author} {\bibinfo {author} {\bibfnamefont {G.}~\bibnamefont
			{Yu}}, \bibinfo {author} {\bibfnamefont {P.}~\bibnamefont {Upadhyaya}},
		\bibinfo {author} {\bibfnamefont {Y.}~\bibnamefont {Fan}}, \bibinfo {author}
		{\bibfnamefont {J.~G.}\ \bibnamefont {Alzate}}, \bibinfo {author}
		{\bibfnamefont {W.}~\bibnamefont {Jiang}}, \bibinfo {author} {\bibfnamefont
			{K.~L.}\ \bibnamefont {Wong}}, \bibinfo {author} {\bibfnamefont
			{S.}~\bibnamefont {Takei}}, \bibinfo {author} {\bibfnamefont {S.~A.}\
			\bibnamefont {Bender}}, \bibinfo {author} {\bibfnamefont {L.-T.}\
			\bibnamefont {Chang}}, \bibinfo {author} {\bibfnamefont {Y.}~\bibnamefont
			{Jiang}}, \bibinfo {author} {\bibfnamefont {M.}~\bibnamefont {Lang}},
		\bibinfo {author} {\bibfnamefont {J.}~\bibnamefont {Tang}}, \bibinfo {author}
		{\bibfnamefont {Y.}~\bibnamefont {Wang}}, \bibinfo {author} {\bibfnamefont
			{Y.}~\bibnamefont {Tserkovnyak}}, \bibinfo {author} {\bibfnamefont {P.~K.}\
			\bibnamefont {Amiri}},\ and\ \bibinfo {author} {\bibfnamefont {K.~L.}\
			\bibnamefont {Wang}},\ }\href {https://doi.org/10.1038/nnano.2014.94}
	{\bibfield  {journal} {\bibinfo  {journal} {Nat. Nanotechnol.}\ }\textbf
		{\bibinfo {volume} {9}},\ \bibinfo {pages} {548} (\bibinfo {year}
		{2014})}\BibitemShut {NoStop}%
	\bibitem [{\citenamefont {Liu}\ \emph {et~al.}(2021)\citenamefont {Liu},
		\citenamefont {Zhou}, \citenamefont {Shu}, \citenamefont {Li}, \citenamefont
		{Zhao}, \citenamefont {Lin}, \citenamefont {Deng}, \citenamefont {Xie},
		\citenamefont {Chen}, \citenamefont {Zhou}, \citenamefont {Guo},
		\citenamefont {Wang}, \citenamefont {Yu}, \citenamefont {Shi}, \citenamefont
		{Yang}, \citenamefont {Pennycook}, \citenamefont {Manchon},\ and\
		\citenamefont {Chen}}]{Liu2021Jan}%
	\BibitemOpen
	\bibfield  {author} {\bibinfo {author} {\bibfnamefont {L.}~\bibnamefont
			{Liu}}, \bibinfo {author} {\bibfnamefont {C.}~\bibnamefont {Zhou}}, \bibinfo
		{author} {\bibfnamefont {X.}~\bibnamefont {Shu}}, \bibinfo {author}
		{\bibfnamefont {C.}~\bibnamefont {Li}}, \bibinfo {author} {\bibfnamefont
			{T.}~\bibnamefont {Zhao}}, \bibinfo {author} {\bibfnamefont {W.}~\bibnamefont
			{Lin}}, \bibinfo {author} {\bibfnamefont {J.}~\bibnamefont {Deng}}, \bibinfo
		{author} {\bibfnamefont {Q.}~\bibnamefont {Xie}}, \bibinfo {author}
		{\bibfnamefont {S.}~\bibnamefont {Chen}}, \bibinfo {author} {\bibfnamefont
			{J.}~\bibnamefont {Zhou}}, \bibinfo {author} {\bibfnamefont {R.}~\bibnamefont
			{Guo}}, \bibinfo {author} {\bibfnamefont {H.}~\bibnamefont {Wang}}, \bibinfo
		{author} {\bibfnamefont {J.}~\bibnamefont {Yu}}, \bibinfo {author}
		{\bibfnamefont {S.}~\bibnamefont {Shi}}, \bibinfo {author} {\bibfnamefont
			{P.}~\bibnamefont {Yang}}, \bibinfo {author} {\bibfnamefont {S.}~\bibnamefont
			{Pennycook}}, \bibinfo {author} {\bibfnamefont {A.}~\bibnamefont {Manchon}},\
		and\ \bibinfo {author} {\bibfnamefont {J.}~\bibnamefont {Chen}},\ }\href
	{https://doi.org/10.1038/s41565-020-00826-8} {\bibfield  {journal} {\bibinfo
			{journal} {Nat. Nanotechnol.}\ }\textbf {\bibinfo {volume} {16}},\ \bibinfo
		{pages} {277} (\bibinfo {year} {2021})}\BibitemShut {NoStop}%
	\bibitem [{\citenamefont {MacNeill}\ \emph {et~al.}(2017)\citenamefont
		{MacNeill}, \citenamefont {Stiehl}, \citenamefont {Guimaraes}, \citenamefont
		{Buhrman}, \citenamefont {Park},\ and\ \citenamefont
		{Ralph}}]{MacNeill2017Mar}%
	\BibitemOpen
	\bibfield  {author} {\bibinfo {author} {\bibfnamefont {D.}~\bibnamefont
			{MacNeill}}, \bibinfo {author} {\bibfnamefont {G.~M.}\ \bibnamefont
			{Stiehl}}, \bibinfo {author} {\bibfnamefont {M.~H.~D.}\ \bibnamefont
			{Guimaraes}}, \bibinfo {author} {\bibfnamefont {R.~A.}\ \bibnamefont
			{Buhrman}}, \bibinfo {author} {\bibfnamefont {J.}~\bibnamefont {Park}},\ and\
		\bibinfo {author} {\bibfnamefont {D.~C.}\ \bibnamefont {Ralph}},\ }\href
	{https://doi.org/10.1038/nphys3933} {\bibfield  {journal} {\bibinfo
			{journal} {Nat. Phys.}\ }\textbf {\bibinfo {volume} {13}},\ \bibinfo {pages}
		{300} (\bibinfo {year} {2017})}\BibitemShut {NoStop}%
	\bibitem [{\citenamefont {Garcia}\ \emph {et~al.}(2020)\citenamefont {Garcia},
		\citenamefont {Vila}, \citenamefont {Hsu}, \citenamefont {Waintal},
		\citenamefont {Pereira},\ and\ \citenamefont
		{Roche}}]{PhysRevLett.125.256603}%
	\BibitemOpen
	\bibfield  {author} {\bibinfo {author} {\bibfnamefont {J.~H.}\ \bibnamefont
			{Garcia}}, \bibinfo {author} {\bibfnamefont {M.}~\bibnamefont {Vila}},
		\bibinfo {author} {\bibfnamefont {C.-H.}\ \bibnamefont {Hsu}}, \bibinfo
		{author} {\bibfnamefont {X.}~\bibnamefont {Waintal}}, \bibinfo {author}
		{\bibfnamefont {V.~M.}\ \bibnamefont {Pereira}},\ and\ \bibinfo {author}
		{\bibfnamefont {S.}~\bibnamefont {Roche}},\ }\href
	{https://doi.org/10.1103/PhysRevLett.125.256603} {\bibfield  {journal}
		{\bibinfo  {journal} {Phys. Rev. Lett.}\ }\textbf {\bibinfo {volume} {125}},\
		\bibinfo {pages} {256603} (\bibinfo {year} {2020})}\BibitemShut {NoStop}%
	\bibitem [{\citenamefont {Taniguchi}\ \emph {et~al.}(2015)\citenamefont
		{Taniguchi}, \citenamefont {Grollier},\ and\ \citenamefont
		{Stiles}}]{PhysRevApplied.3.044001}%
	\BibitemOpen
	\bibfield  {author} {\bibinfo {author} {\bibfnamefont {T.}~\bibnamefont
			{Taniguchi}}, \bibinfo {author} {\bibfnamefont {J.}~\bibnamefont
			{Grollier}},\ and\ \bibinfo {author} {\bibfnamefont {M.~D.}\ \bibnamefont
			{Stiles}},\ }\href {https://doi.org/10.1103/PhysRevApplied.3.044001}
	{\bibfield  {journal} {\bibinfo  {journal} {Phys. Rev. Applied}\ }\textbf
		{\bibinfo {volume} {3}},\ \bibinfo {pages} {044001} (\bibinfo {year}
		{2015})}\BibitemShut {NoStop}%
	\bibitem [{\citenamefont {Koike}\ \emph {et~al.}(2020)\citenamefont {Koike},
		\citenamefont {Iihama},\ and\ \citenamefont {Mizukami}}]{Koike2020Aug}%
	\BibitemOpen
	\bibfield  {author} {\bibinfo {author} {\bibfnamefont {Y.}~\bibnamefont
			{Koike}}, \bibinfo {author} {\bibfnamefont {S.}~\bibnamefont {Iihama}},\ and\
		\bibinfo {author} {\bibfnamefont {S.}~\bibnamefont {Mizukami}},\ }\href
	{https://doi.org/10.35848/1347-4065/abac40} {\bibfield  {journal} {\bibinfo
			{journal} {Jpn. J. Appl. Phys.}\ }\textbf {\bibinfo {volume} {59}},\ \bibinfo
		{pages} {090907} (\bibinfo {year} {2020})}\BibitemShut {NoStop}%
	\bibitem [{\citenamefont {Varotto}\ \emph {et~al.}(2020)\citenamefont
		{Varotto}, \citenamefont {Cosset-Ch\'eneau}, \citenamefont {Gr\`ezes},
		\citenamefont {Fu}, \citenamefont {Warin}, \citenamefont {Brenac},
		\citenamefont {Jacquot}, \citenamefont {Gambarelli}, \citenamefont {Rinaldi},
		\citenamefont {Baltz}, \citenamefont {Attan\'e}, \citenamefont {Vila},\ and\
		\citenamefont {No\"el}}]{PhysRevLett.125.267204}%
	\BibitemOpen
	\bibfield  {author} {\bibinfo {author} {\bibfnamefont {S.}~\bibnamefont
			{Varotto}}, \bibinfo {author} {\bibfnamefont {M.}~\bibnamefont
			{Cosset-Ch\'eneau}}, \bibinfo {author} {\bibfnamefont {C.}~\bibnamefont
			{Gr\`ezes}}, \bibinfo {author} {\bibfnamefont {Y.}~\bibnamefont {Fu}},
		\bibinfo {author} {\bibfnamefont {P.}~\bibnamefont {Warin}}, \bibinfo
		{author} {\bibfnamefont {A.}~\bibnamefont {Brenac}}, \bibinfo {author}
		{\bibfnamefont {J.F.}~\bibnamefont {Jacquot}}, \bibinfo {author} {\bibfnamefont
			{S.}~\bibnamefont {Gambarelli}}, \bibinfo {author} {\bibfnamefont
			{C.}~\bibnamefont {Rinaldi}}, \bibinfo {author} {\bibfnamefont
			{V.}~\bibnamefont {Baltz}}, \bibinfo {author} {\bibfnamefont {J.-P.}\
			\bibnamefont {Attan\'e}}, \bibinfo {author} {\bibfnamefont {L.}~\bibnamefont
			{Vila}},\ and\ \bibinfo {author} {\bibfnamefont {P.}~\bibnamefont {No\"el}},\
	}\href {https://doi.org/10.1103/PhysRevLett.125.267204} {\bibfield  {journal}
		{\bibinfo  {journal} {Phys. Rev. Lett.}\ }\textbf {\bibinfo {volume} {125}},\
		\bibinfo {pages} {267204} (\bibinfo {year} {2020})}\BibitemShut {NoStop}%
	\bibitem [{\citenamefont {Cramer}\ \emph {et~al.}(2019)\citenamefont {Cramer},
		\citenamefont {Ross}, \citenamefont {Jaiswal}, \citenamefont {Baldrati},
		\citenamefont {Lebrun},\ and\ \citenamefont {Kl\"aui}}]{PhysRevB.99.104414}%
	\BibitemOpen
	\bibfield  {author} {\bibinfo {author} {\bibfnamefont {J.}~\bibnamefont
			{Cramer}}, \bibinfo {author} {\bibfnamefont {A.}~\bibnamefont {Ross}},
		\bibinfo {author} {\bibfnamefont {S.}~\bibnamefont {Jaiswal}}, \bibinfo
		{author} {\bibfnamefont {L.}~\bibnamefont {Baldrati}}, \bibinfo {author}
		{\bibfnamefont {R.}~\bibnamefont {Lebrun}},\ and\ \bibinfo {author}
		{\bibfnamefont {M.}~\bibnamefont {Kl\"aui}},\ }\href
	{https://doi.org/10.1103/PhysRevB.99.104414} {\bibfield  {journal} {\bibinfo
			{journal} {Phys. Rev. B}\ }\textbf {\bibinfo {volume} {99}},\ \bibinfo
		{pages} {104414} (\bibinfo {year} {2019})}\BibitemShut {NoStop}%
	\bibitem [{\citenamefont {Nagaosa}\ \emph {et~al.}(2010)\citenamefont
		{Nagaosa}, \citenamefont {Sinova}, \citenamefont {Onoda}, \citenamefont
		{MacDonald},\ and\ \citenamefont {Ong}}]{RevModPhys.82.1539}%
	\BibitemOpen
	\bibfield  {author} {\bibinfo {author} {\bibfnamefont {N.}~\bibnamefont
			{Nagaosa}}, \bibinfo {author} {\bibfnamefont {J.}~\bibnamefont {Sinova}},
		\bibinfo {author} {\bibfnamefont {S.}~\bibnamefont {Onoda}}, \bibinfo
		{author} {\bibfnamefont {A.~H.}\ \bibnamefont {MacDonald}},\ and\ \bibinfo
		{author} {\bibfnamefont {N.~P.}\ \bibnamefont {Ong}},\ }\href
	{https://doi.org/10.1103/RevModPhys.82.1539} {\bibfield  {journal} {\bibinfo
			{journal} {Rev. Mod. Phys.}\ }\textbf {\bibinfo {volume} {82}},\ \bibinfo
		{pages} {1539} (\bibinfo {year} {2010})}\BibitemShut {NoStop}%
	\bibitem [{\citenamefont {Omori}\ \emph {et~al.}(2019)\citenamefont {Omori},
		\citenamefont {Sagasta}, \citenamefont {Niimi}, \citenamefont {Gradhand},
		\citenamefont {Hueso}, \citenamefont {Casanova},\ and\ \citenamefont
		{Otani}}]{PhysRevB.99.014403}%
	\BibitemOpen
	\bibfield  {author} {\bibinfo {author} {\bibfnamefont {Y.}~\bibnamefont
			{Omori}}, \bibinfo {author} {\bibfnamefont {E.}~\bibnamefont {Sagasta}},
		\bibinfo {author} {\bibfnamefont {Y.}~\bibnamefont {Niimi}}, \bibinfo
		{author} {\bibfnamefont {M.}~\bibnamefont {Gradhand}}, \bibinfo {author}
		{\bibfnamefont {L.~E.}\ \bibnamefont {Hueso}}, \bibinfo {author}
		{\bibfnamefont {F.}~\bibnamefont {Casanova}},\ and\ \bibinfo {author}
		{\bibfnamefont {Y.~C.}\ \bibnamefont {Otani}},\ }\href
	{https://doi.org/10.1103/PhysRevB.99.014403} {\bibfield  {journal} {\bibinfo
			{journal} {Phys. Rev. B}\ }\textbf {\bibinfo {volume} {99}},\ \bibinfo
		{pages} {014403} (\bibinfo {year} {2019})}\BibitemShut {NoStop}%
	\bibitem [{\citenamefont {Iihama}\ \emph {et~al.}(2018)\citenamefont {Iihama},
		\citenamefont {Taniguchi}, \citenamefont {Yakushiji}, \citenamefont
		{Fukushima}, \citenamefont {Shiota}, \citenamefont {Tsunegi}, \citenamefont
		{Hiramatsu}, \citenamefont {Yuasa}, \citenamefont {Suzuki},\ and\
		\citenamefont {Kubota}}]{Iihama2018Feb}%
	\BibitemOpen
	\bibfield  {author} {\bibinfo {author} {\bibfnamefont {S.}~\bibnamefont
			{Iihama}}, \bibinfo {author} {\bibfnamefont {T.}~\bibnamefont {Taniguchi}},
		\bibinfo {author} {\bibfnamefont {K.}~\bibnamefont {Yakushiji}}, \bibinfo
		{author} {\bibfnamefont {A.}~\bibnamefont {Fukushima}}, \bibinfo {author}
		{\bibfnamefont {Y.}~\bibnamefont {Shiota}}, \bibinfo {author} {\bibfnamefont
			{S.}~\bibnamefont {Tsunegi}}, \bibinfo {author} {\bibfnamefont
			{R.}~\bibnamefont {Hiramatsu}}, \bibinfo {author} {\bibfnamefont
			{S.}~\bibnamefont {Yuasa}}, \bibinfo {author} {\bibfnamefont
			{Y.}~\bibnamefont {Suzuki}},\ and\ \bibinfo {author} {\bibfnamefont
			{H.}~\bibnamefont {Kubota}},\ }\href
	{https://doi.org/10.1038/s41928-018-0026-z} {\bibfield  {journal} {\bibinfo
			{journal} {Nat. Electron.}\ }\textbf {\bibinfo {volume} {1}},\ \bibinfo
		{pages} {120} (\bibinfo {year} {2018})}\BibitemShut {NoStop}%
	\bibitem [{\citenamefont {Seki}\ \emph {et~al.}(2019)\citenamefont {Seki},
		\citenamefont {Iihama}, \citenamefont {Taniguchi},\ and\ \citenamefont
		{Takanashi}}]{PhysRevB.100.144427}%
	\BibitemOpen
	\bibfield  {author} {\bibinfo {author} {\bibfnamefont {T.}~\bibnamefont
			{Seki}}, \bibinfo {author} {\bibfnamefont {S.}~\bibnamefont {Iihama}},
		\bibinfo {author} {\bibfnamefont {T.}~\bibnamefont {Taniguchi}},\ and\
		\bibinfo {author} {\bibfnamefont {K.}~\bibnamefont {Takanashi}},\ }\href
	{https://doi.org/10.1103/PhysRevB.100.144427} {\bibfield  {journal} {\bibinfo
			{journal} {Phys. Rev. B}\ }\textbf {\bibinfo {volume} {100}},\ \bibinfo
		{pages} {144427} (\bibinfo {year} {2019})}\BibitemShut {NoStop}%
	\bibitem [{\citenamefont {Chuang}\ \emph {et~al.}(2020)\citenamefont {Chuang},
		\citenamefont {Qu}, \citenamefont {Huang},\ and\ \citenamefont
		{Lee}}]{PhysRevResearch.2.032053}%
	\BibitemOpen
	\bibfield  {author} {\bibinfo {author} {\bibfnamefont {T.~C.}\ \bibnamefont
			{Chuang}}, \bibinfo {author} {\bibfnamefont {D.}~\bibnamefont {Qu}}, \bibinfo
		{author} {\bibfnamefont {S.~Y.}\ \bibnamefont {Huang}},\ and\ \bibinfo
		{author} {\bibfnamefont {S.~F.}\ \bibnamefont {Lee}},\ }\href
	{https://doi.org/10.1103/PhysRevResearch.2.032053} {\bibfield  {journal}
		{\bibinfo  {journal} {Phys. Rev. Research}\ }\textbf {\bibinfo {volume}
			{2}},\ \bibinfo {pages} {032053(R)} (\bibinfo {year} {2020})}\BibitemShut
	{NoStop}%
	\bibitem [{\citenamefont {Amin}\ \emph {et~al.}(2019)\citenamefont {Amin},
		\citenamefont {Li}, \citenamefont {Stiles},\ and\ \citenamefont
		{Haney}}]{PhysRevB.99.220405}%
	\BibitemOpen
	\bibfield  {author} {\bibinfo {author} {\bibfnamefont {V.~P.}\ \bibnamefont
			{Amin}}, \bibinfo {author} {\bibfnamefont {J.}~\bibnamefont {Li}}, \bibinfo
		{author} {\bibfnamefont {M.~D.}\ \bibnamefont {Stiles}},\ and\ \bibinfo
		{author} {\bibfnamefont {P.~M.}\ \bibnamefont {Haney}},\ }\href
	{https://doi.org/10.1103/PhysRevB.99.220405} {\bibfield  {journal} {\bibinfo
			{journal} {Phys. Rev. B}\ }\textbf {\bibinfo {volume} {99}},\ \bibinfo
		{pages} {220405(R)} (\bibinfo {year} {2019})}\BibitemShut {NoStop}%
	\bibitem [{\citenamefont {Davidson}\ \emph {et~al.}(2020)\citenamefont
		{Davidson}, \citenamefont {Amin}, \citenamefont {Aljuaid}, \citenamefont
		{Haney},\ and\ \citenamefont {Fan}}]{DAVIDSON2020126228}%
	\BibitemOpen
	\bibfield  {author} {\bibinfo {author} {\bibfnamefont {A.}~\bibnamefont
			{Davidson}}, \bibinfo {author} {\bibfnamefont {V.~P.}\ \bibnamefont {Amin}},
		\bibinfo {author} {\bibfnamefont {W.~S.}\ \bibnamefont {Aljuaid}}, \bibinfo
		{author} {\bibfnamefont {P.~M.}\ \bibnamefont {Haney}},\ and\ \bibinfo
		{author} {\bibfnamefont {X.}~\bibnamefont {Fan}},\ }\href
	{https://doi.org/https://doi.org/10.1016/j.physleta.2019.126228} {\bibfield
		{journal} {\bibinfo  {journal} {Phys. Lett. A}\ }\textbf {\bibinfo {volume}
			{384}},\ \bibinfo {pages} {126228} (\bibinfo {year} {2020})}\BibitemShut
	{NoStop}%
	\bibitem [{\citenamefont {Qu}\ \emph {et~al.}(2020)\citenamefont {Qu},
		\citenamefont {Nakamura},\ and\ \citenamefont
		{Hayashi}}]{PhysRevB.102.144440}%
	\BibitemOpen
	\bibfield  {author} {\bibinfo {author} {\bibfnamefont {G.}~\bibnamefont
			{Qu}}, \bibinfo {author} {\bibfnamefont {K.}~\bibnamefont {Nakamura}},\ and\
		\bibinfo {author} {\bibfnamefont {M.}~\bibnamefont {Hayashi}},\ }\href
	{https://doi.org/10.1103/PhysRevB.102.144440} {\bibfield  {journal} {\bibinfo
			{journal} {Phys. Rev. B}\ }\textbf {\bibinfo {volume} {102}},\ \bibinfo
		{pages} {144440} (\bibinfo {year} {2020})}\BibitemShut {NoStop}%
	\bibitem [{\citenamefont {Burch}\ \emph {et~al.}(2018)\citenamefont {Burch},
		\citenamefont {Mandrus},\ and\ \citenamefont {Park}}]{Burch2018Nov}%
	\BibitemOpen
	\bibfield  {author} {\bibinfo {author} {\bibfnamefont {K.~S.}\ \bibnamefont
			{Burch}}, \bibinfo {author} {\bibfnamefont {D.}~\bibnamefont {Mandrus}},\
		and\ \bibinfo {author} {\bibfnamefont {J.-G.}\ \bibnamefont {Park}},\ }\href
	{https://doi.org/10.1038/s41586-018-0631-z} {\bibfield  {journal} {\bibinfo
			{journal} {Nature}\ }\textbf {\bibinfo {volume} {563}},\ \bibinfo {pages}
		{47} (\bibinfo {year} {2018})}\BibitemShut {NoStop}%
	\bibitem [{\citenamefont {Lin}\ \emph {et~al.}(2019)\citenamefont {Lin},
		\citenamefont {Yang}, \citenamefont {Wang},\ and\ \citenamefont
		{Zhao}}]{Lin2019Jul}%
	\BibitemOpen
	\bibfield  {author} {\bibinfo {author} {\bibfnamefont {X.}~\bibnamefont
			{Lin}}, \bibinfo {author} {\bibfnamefont {W.}~\bibnamefont {Yang}}, \bibinfo
		{author} {\bibfnamefont {K.~L.}\ \bibnamefont {Wang}},\ and\ \bibinfo
		{author} {\bibfnamefont {W.}~\bibnamefont {Zhao}},\ }\href
	{https://doi.org/10.1038/s41928-019-0273-7} {\bibfield  {journal} {\bibinfo
			{journal} {Nat. Electron.}\ }\textbf {\bibinfo {volume} {2}},\ \bibinfo
		{pages} {274} (\bibinfo {year} {2019})}\BibitemShut {NoStop}%
	\bibitem [{\citenamefont {Fei}\ \emph {et~al.}(2018)\citenamefont {Fei},
		\citenamefont {Huang}, \citenamefont {Malinowski}, \citenamefont {Wang},
		\citenamefont {Song}, \citenamefont {Sanchez}, \citenamefont {Yao},
		\citenamefont {Xiao}, \citenamefont {Zhu}, \citenamefont {May}, \citenamefont
		{Wu}, \citenamefont {Cobden}, \citenamefont {Chu},\ and\ \citenamefont
		{Xu}}]{Fei2018Sep}%
	\BibitemOpen
	\bibfield  {author} {\bibinfo {author} {\bibfnamefont {Z.}~\bibnamefont
			{Fei}}, \bibinfo {author} {\bibfnamefont {B.}~\bibnamefont {Huang}}, \bibinfo
		{author} {\bibfnamefont {P.}~\bibnamefont {Malinowski}}, \bibinfo {author}
		{\bibfnamefont {W.}~\bibnamefont {Wang}}, \bibinfo {author} {\bibfnamefont
			{T.}~\bibnamefont {Song}}, \bibinfo {author} {\bibfnamefont {J.}~\bibnamefont
			{Sanchez}}, \bibinfo {author} {\bibfnamefont {W.}~\bibnamefont {Yao}},
		\bibinfo {author} {\bibfnamefont {D.}~\bibnamefont {Xiao}}, \bibinfo {author}
		{\bibfnamefont {X.}~\bibnamefont {Zhu}}, \bibinfo {author} {\bibfnamefont
			{A.~F.}\ \bibnamefont {May}}, \bibinfo {author} {\bibfnamefont
			{W.}~\bibnamefont {Wu}}, \bibinfo {author} {\bibfnamefont {D.~H.}\
			\bibnamefont {Cobden}}, \bibinfo {author} {\bibfnamefont {J.-H.}\
			\bibnamefont {Chu}},\ and\ \bibinfo {author} {\bibfnamefont {X.}~\bibnamefont
			{Xu}},\ }\href {https://doi.org/10.1038/s41563-018-0149-7} {\bibfield
		{journal} {\bibinfo  {journal} {Nat. Mater.}\ }\textbf {\bibinfo {volume}
			{17}},\ \bibinfo {pages} {778} (\bibinfo {year} {2018})}\BibitemShut
	{NoStop}%
	\bibitem [{\citenamefont {Deng}\ \emph {et~al.}(2018)\citenamefont {Deng},
		\citenamefont {Yu}, \citenamefont {Song}, \citenamefont {Zhang},
		\citenamefont {Wang}, \citenamefont {Sun}, \citenamefont {Yi}, \citenamefont
		{Wu}, \citenamefont {Wu}, \citenamefont {Zhu}, \citenamefont {Wang},
		\citenamefont {Chen},\ and\ \citenamefont {Zhang}}]{Deng2018Nov}%
	\BibitemOpen
	\bibfield  {author} {\bibinfo {author} {\bibfnamefont {Y.}~\bibnamefont
			{Deng}}, \bibinfo {author} {\bibfnamefont {Y.}~\bibnamefont {Yu}}, \bibinfo
		{author} {\bibfnamefont {Y.}~\bibnamefont {Song}}, \bibinfo {author}
		{\bibfnamefont {J.}~\bibnamefont {Zhang}}, \bibinfo {author} {\bibfnamefont
			{N.~Z.}\ \bibnamefont {Wang}}, \bibinfo {author} {\bibfnamefont
			{Z.}~\bibnamefont {Sun}}, \bibinfo {author} {\bibfnamefont {Y.}~\bibnamefont
			{Yi}}, \bibinfo {author} {\bibfnamefont {Y.~Z.}\ \bibnamefont {Wu}}, \bibinfo
		{author} {\bibfnamefont {S.}~\bibnamefont {Wu}}, \bibinfo {author}
		{\bibfnamefont {J.}~\bibnamefont {Zhu}}, \bibinfo {author} {\bibfnamefont
			{J.}~\bibnamefont {Wang}}, \bibinfo {author} {\bibfnamefont {X.~H.}\
			\bibnamefont {Chen}},\ and\ \bibinfo {author} {\bibfnamefont
			{Y.}~\bibnamefont {Zhang}},\ }\href
	{https://doi.org/10.1038/s41586-018-0626-9} {\bibfield  {journal} {\bibinfo
			{journal} {Nature}\ }\textbf {\bibinfo {volume} {563}},\ \bibinfo {pages}
		{94} (\bibinfo {year} {2018})}\BibitemShut {NoStop}%
	\bibitem [{\citenamefont {Newnham}(2005)}]{Newnham2005Jan}%
	\BibitemOpen
	\bibfield  {author} {\bibinfo {author} {\bibfnamefont {R.~E.}\ \bibnamefont
			{Newnham}},\ }\href@noop {} {\emph {\bibinfo {title} {{Properties of
					Materials: Anisotropy, Symmetry, Structure}}}}\ (\bibinfo  {publisher}
	{Oxford University Press},\ \bibinfo {address} {Oxford, England, UK},\
	\bibinfo {year} {2005})\BibitemShut {NoStop}%
	\bibitem [{\citenamefont {Gallego}\ \emph {et~al.}(2019)\citenamefont
		{Gallego}, \citenamefont {Etxebarria}, \citenamefont {Elcoro}, \citenamefont
		{Tasci},\ and\ \citenamefont {Perez-Mato}}]{Gallegolk5043}%
	\BibitemOpen
	\bibfield  {author} {\bibinfo {author} {\bibfnamefont {S.~V.}\ \bibnamefont
			{Gallego}}, \bibinfo {author} {\bibfnamefont {J.}~\bibnamefont {Etxebarria}},
		\bibinfo {author} {\bibfnamefont {L.}~\bibnamefont {Elcoro}}, \bibinfo
		{author} {\bibfnamefont {E.~S.}\ \bibnamefont {Tasci}},\ and\ \bibinfo
		{author} {\bibfnamefont {J.~M.}\ \bibnamefont {Perez-Mato}},\ }\href
	{https://doi.org/10.1107/S2053273319001748} {\bibfield  {journal} {\bibinfo
			{journal} {Acta Crystallogr., Sect. A}\ }\textbf {\bibinfo {volume} {75}},\
		\bibinfo {pages} {438} (\bibinfo {year} {2019})}\BibitemShut {NoStop}%
	\bibitem [{\citenamefont {Seemann}\ \emph {et~al.}(2015)\citenamefont
		{Seemann}, \citenamefont {K\"odderitzsch}, \citenamefont {Wimmer},\ and\
		\citenamefont {Ebert}}]{PhysRevB.92.155138}%
	\BibitemOpen
	\bibfield  {author} {\bibinfo {author} {\bibfnamefont {M.}~\bibnamefont
			{Seemann}}, \bibinfo {author} {\bibfnamefont {D.}~\bibnamefont
			{K\"odderitzsch}}, \bibinfo {author} {\bibfnamefont {S.}~\bibnamefont
			{Wimmer}},\ and\ \bibinfo {author} {\bibfnamefont {H.}~\bibnamefont
			{Ebert}},\ }\href {https://doi.org/10.1103/PhysRevB.92.155138} {\bibfield
		{journal} {\bibinfo  {journal} {Phys. Rev. B}\ }\textbf {\bibinfo {volume}
			{92}},\ \bibinfo {pages} {155138} (\bibinfo {year} {2015})}\BibitemShut
	{NoStop}%
	\bibitem [{\citenamefont {Freimuth}\ \emph {et~al.}(2021)\citenamefont
		{Freimuth}, \citenamefont {Bl{\ifmmode\ddot{u}\else\"{u}\fi}gel},\ and\
		\citenamefont {Mokrousov}}]{Freimuth2021Mar}%
	\BibitemOpen
	\bibfield  {author} {\bibinfo {author} {\bibfnamefont {F.}~\bibnamefont
			{Freimuth}}, \bibinfo {author} {\bibfnamefont {S.}~\bibnamefont
			{Bl{\ifmmode\ddot{u}\else\"{u}\fi}gel}},\ and\ \bibinfo {author}
		{\bibfnamefont {Y.}~\bibnamefont {Mokrousov}},\ }\href
	{https://arxiv.org/abs/2103.15663v1} {\bibfield  {journal} {\bibinfo
			{journal} {arXiv}\ } (\bibinfo {year} {2021})},\ \Eprint
	{https://arxiv.org/abs/2103.15663} {2103.15663} \BibitemShut {NoStop}%
	\bibitem [{\citenamefont {Freimuth}\ \emph {et~al.}(2010)\citenamefont
		{Freimuth}, \citenamefont {Bl\"ugel},\ and\ \citenamefont
		{Mokrousov}}]{PhysRevLett.105.246602}%
	\BibitemOpen
	\bibfield  {author} {\bibinfo {author} {\bibfnamefont {F.}~\bibnamefont
			{Freimuth}}, \bibinfo {author} {\bibfnamefont {S.}~\bibnamefont {Bl\"ugel}},\
		and\ \bibinfo {author} {\bibfnamefont {Y.}~\bibnamefont {Mokrousov}},\ }\href
	{https://doi.org/10.1103/PhysRevLett.105.246602} {\bibfield  {journal}
		{\bibinfo  {journal} {Phys. Rev. Lett.}\ }\textbf {\bibinfo {volume} {105}},\
		\bibinfo {pages} {246602} (\bibinfo {year} {2010})}\BibitemShut {NoStop}%
	\bibitem [{\citenamefont {Giannozzi}\ \emph {et~al.}(2017)\citenamefont
		{Giannozzi}, \citenamefont {Andreussi}, \citenamefont {Brumme}, \citenamefont
		{Bunau}, \citenamefont {Nardelli}, \citenamefont {Calandra}, \citenamefont
		{Car}, \citenamefont {Cavazzoni}, \citenamefont {Ceresoli}, \citenamefont
		{Cococcioni}, \citenamefont {Colonna}, \citenamefont {Carnimeo},
		\citenamefont {Corso}, \citenamefont {de~Gironcoli}, \citenamefont {Delugas},
		\citenamefont {DiStasio}, \citenamefont {Ferretti}, \citenamefont {Floris},
		\citenamefont {Fratesi}, \citenamefont {Fugallo}, \citenamefont {Gebauer},
		\citenamefont {Gerstmann}, \citenamefont {Giustino}, \citenamefont {Gorni},
		\citenamefont {Jia}, \citenamefont {Kawamura}, \citenamefont {Ko},
		\citenamefont {Kokalj}, \citenamefont {Kü{\c{c}}ükbenli}, \citenamefont
		{Lazzeri}, \citenamefont {Marsili}, \citenamefont {Marzari}, \citenamefont
		{Mauri}, \citenamefont {Nguyen}, \citenamefont {Nguyen}, \citenamefont {de-la
			Roza}, \citenamefont {Paulatto}, \citenamefont {Ponc{\'{e}}}, \citenamefont
		{Rocca}, \citenamefont {Sabatini}, \citenamefont {Santra}, \citenamefont
		{Schlipf}, \citenamefont {Seitsonen}, \citenamefont {Smogunov}, \citenamefont
		{Timrov}, \citenamefont {Thonhauser}, \citenamefont {Umari}, \citenamefont
		{Vast}, \citenamefont {Wu},\ and\ \citenamefont {Baroni}}]{Giannozzi_2017}%
	\BibitemOpen
	\bibfield  {author} {\bibinfo {author} {\bibfnamefont {P.}~\bibnamefont
			{Giannozzi}}, \bibinfo {author} {\bibfnamefont {O.}~\bibnamefont
			{Andreussi}}, \bibinfo {author} {\bibfnamefont {T.}~\bibnamefont {Brumme}},
		\bibinfo {author} {\bibfnamefont {O.}~\bibnamefont {Bunau}}, \bibinfo
		{author} {\bibfnamefont {M.~B.}\ \bibnamefont {Nardelli}}, \bibinfo {author}
		{\bibfnamefont {M.}~\bibnamefont {Calandra}}, \bibinfo {author}
		{\bibfnamefont {R.}~\bibnamefont {Car}}, \bibinfo {author} {\bibfnamefont
			{C.}~\bibnamefont {Cavazzoni}}, \bibinfo {author} {\bibfnamefont
			{D.}~\bibnamefont {Ceresoli}}, \bibinfo {author} {\bibfnamefont
			{M.}~\bibnamefont {Cococcioni}}, \bibinfo {author} {\bibfnamefont
			{N.}~\bibnamefont {Colonna}}, \bibinfo {author} {\bibfnamefont
			{I.}~\bibnamefont {Carnimeo}}, \bibinfo {author} {\bibfnamefont {A.~D.}\
			\bibnamefont {Corso}}, \bibinfo {author} {\bibfnamefont {S.}~\bibnamefont
			{de~Gironcoli}}, \bibinfo {author} {\bibfnamefont {P.}~\bibnamefont
			{Delugas}}, \bibinfo {author} {\bibfnamefont {R.~A.}\ \bibnamefont
			{DiStasio}}, \bibinfo {author} {\bibfnamefont {A.}~\bibnamefont {Ferretti}},
		\bibinfo {author} {\bibfnamefont {A.}~\bibnamefont {Floris}}, \bibinfo
		{author} {\bibfnamefont {G.}~\bibnamefont {Fratesi}}, \bibinfo {author}
		{\bibfnamefont {G.}~\bibnamefont {Fugallo}}, \bibinfo {author} {\bibfnamefont
			{R.}~\bibnamefont {Gebauer}}, \bibinfo {author} {\bibfnamefont
			{U.}~\bibnamefont {Gerstmann}}, \bibinfo {author} {\bibfnamefont
			{F.}~\bibnamefont {Giustino}}, \bibinfo {author} {\bibfnamefont
			{T.}~\bibnamefont {Gorni}}, \bibinfo {author} {\bibfnamefont
			{J.}~\bibnamefont {Jia}}, \bibinfo {author} {\bibfnamefont {M.}~\bibnamefont
			{Kawamura}}, \bibinfo {author} {\bibfnamefont {H.-Y.}\ \bibnamefont {Ko}},
		\bibinfo {author} {\bibfnamefont {A.}~\bibnamefont {Kokalj}}, \bibinfo
		{author} {\bibfnamefont {E.}~\bibnamefont {Kü{\c{c}}ükbenli}}, \bibinfo
		{author} {\bibfnamefont {M.}~\bibnamefont {Lazzeri}}, \bibinfo {author}
		{\bibfnamefont {M.}~\bibnamefont {Marsili}}, \bibinfo {author} {\bibfnamefont
			{N.}~\bibnamefont {Marzari}}, \bibinfo {author} {\bibfnamefont
			{F.}~\bibnamefont {Mauri}}, \bibinfo {author} {\bibfnamefont {N.~L.}\
			\bibnamefont {Nguyen}}, \bibinfo {author} {\bibfnamefont {H.-V.}\
			\bibnamefont {Nguyen}}, \bibinfo {author} {\bibfnamefont {A.~O.}\
			\bibnamefont {de-la Roza}}, \bibinfo {author} {\bibfnamefont
			{L.}~\bibnamefont {Paulatto}}, \bibinfo {author} {\bibfnamefont
			{S.}~\bibnamefont {Ponc{\'{e}}}}, \bibinfo {author} {\bibfnamefont
			{D.}~\bibnamefont {Rocca}}, \bibinfo {author} {\bibfnamefont
			{R.}~\bibnamefont {Sabatini}}, \bibinfo {author} {\bibfnamefont
			{B.}~\bibnamefont {Santra}}, \bibinfo {author} {\bibfnamefont
			{M.}~\bibnamefont {Schlipf}}, \bibinfo {author} {\bibfnamefont {A.~P.}\
			\bibnamefont {Seitsonen}}, \bibinfo {author} {\bibfnamefont {A.}~\bibnamefont
			{Smogunov}}, \bibinfo {author} {\bibfnamefont {I.}~\bibnamefont {Timrov}},
		\bibinfo {author} {\bibfnamefont {T.}~\bibnamefont {Thonhauser}}, \bibinfo
		{author} {\bibfnamefont {P.}~\bibnamefont {Umari}}, \bibinfo {author}
		{\bibfnamefont {N.}~\bibnamefont {Vast}}, \bibinfo {author} {\bibfnamefont
			{X.}~\bibnamefont {Wu}},\ and\ \bibinfo {author} {\bibfnamefont
			{S.}~\bibnamefont {Baroni}},\ }\href
	{https://doi.org/10.1088/1361-648x/aa8f79} {\bibfield  {journal} {\bibinfo
			{journal} {J. Phys.: Condens. Matter}\ }\textbf {\bibinfo {volume} {29}},\
		\bibinfo {pages} {465901} (\bibinfo {year} {2017})}\BibitemShut {NoStop}%
	\bibitem [{\citenamefont {Pizzi}\ \emph {et~al.}(2020)\citenamefont {Pizzi},
		\citenamefont {Vitale}, \citenamefont {Arita}, \citenamefont {Blügel},
		\citenamefont {Freimuth}, \citenamefont {G{\'{e}}ranton}, \citenamefont
		{Gibertini}, \citenamefont {Gresch}, \citenamefont {Johnson}, \citenamefont
		{Koretsune}, \citenamefont {Iba{\~{n}}ez-Azpiroz}, \citenamefont {Lee},
		\citenamefont {Lihm}, \citenamefont {Marchand}, \citenamefont {Marrazzo},
		\citenamefont {Mokrousov}, \citenamefont {Mustafa}, \citenamefont {Nohara},
		\citenamefont {Nomura}, \citenamefont {Paulatto}, \citenamefont
		{Ponc{\'{e}}}, \citenamefont {Ponweiser}, \citenamefont {Qiao}, \citenamefont
		{Thöle}, \citenamefont {Tsirkin}, \citenamefont {Wierzbowska}, \citenamefont
		{Marzari}, \citenamefont {Vanderbilt}, \citenamefont {Souza}, \citenamefont
		{Mostofi},\ and\ \citenamefont {Yates}}]{Pizzi_2020}%
	\BibitemOpen
	\bibfield  {author} {\bibinfo {author} {\bibfnamefont {G.}~\bibnamefont
			{Pizzi}}, \bibinfo {author} {\bibfnamefont {V.}~\bibnamefont {Vitale}},
		\bibinfo {author} {\bibfnamefont {R.}~\bibnamefont {Arita}}, \bibinfo
		{author} {\bibfnamefont {S.}~\bibnamefont {Blügel}}, \bibinfo {author}
		{\bibfnamefont {F.}~\bibnamefont {Freimuth}}, \bibinfo {author}
		{\bibfnamefont {G.}~\bibnamefont {G{\'{e}}ranton}}, \bibinfo {author}
		{\bibfnamefont {M.}~\bibnamefont {Gibertini}}, \bibinfo {author}
		{\bibfnamefont {D.}~\bibnamefont {Gresch}}, \bibinfo {author} {\bibfnamefont
			{C.}~\bibnamefont {Johnson}}, \bibinfo {author} {\bibfnamefont
			{T.}~\bibnamefont {Koretsune}}, \bibinfo {author} {\bibfnamefont
			{J.}~\bibnamefont {Iba{\~{n}}ez-Azpiroz}}, \bibinfo {author} {\bibfnamefont
			{H.}~\bibnamefont {Lee}}, \bibinfo {author} {\bibfnamefont {J.-M.}\
			\bibnamefont {Lihm}}, \bibinfo {author} {\bibfnamefont {D.}~\bibnamefont
			{Marchand}}, \bibinfo {author} {\bibfnamefont {A.}~\bibnamefont {Marrazzo}},
		\bibinfo {author} {\bibfnamefont {Y.}~\bibnamefont {Mokrousov}}, \bibinfo
		{author} {\bibfnamefont {J.~I.}\ \bibnamefont {Mustafa}}, \bibinfo {author}
		{\bibfnamefont {Y.}~\bibnamefont {Nohara}}, \bibinfo {author} {\bibfnamefont
			{Y.}~\bibnamefont {Nomura}}, \bibinfo {author} {\bibfnamefont
			{L.}~\bibnamefont {Paulatto}}, \bibinfo {author} {\bibfnamefont
			{S.}~\bibnamefont {Ponc{\'{e}}}}, \bibinfo {author} {\bibfnamefont
			{T.}~\bibnamefont {Ponweiser}}, \bibinfo {author} {\bibfnamefont
			{J.}~\bibnamefont {Qiao}}, \bibinfo {author} {\bibfnamefont {F.}~\bibnamefont
			{Thöle}}, \bibinfo {author} {\bibfnamefont {S.~S.}\ \bibnamefont {Tsirkin}},
		\bibinfo {author} {\bibfnamefont {M.}~\bibnamefont {Wierzbowska}}, \bibinfo
		{author} {\bibfnamefont {N.}~\bibnamefont {Marzari}}, \bibinfo {author}
		{\bibfnamefont {D.}~\bibnamefont {Vanderbilt}}, \bibinfo {author}
		{\bibfnamefont {I.}~\bibnamefont {Souza}}, \bibinfo {author} {\bibfnamefont
			{A.~A.}\ \bibnamefont {Mostofi}},\ and\ \bibinfo {author} {\bibfnamefont
			{J.~R.}\ \bibnamefont {Yates}},\ }\href
	{https://doi.org/10.1088/1361-648x/ab51ff} {\bibfield  {journal} {\bibinfo
			{journal} {J. Phys.: Condens. Matter}\ }\textbf {\bibinfo {volume} {32}},\
		\bibinfo {pages} {165902} (\bibinfo {year} {2020})}\BibitemShut {NoStop}%
	\bibitem [{sup()}]{supp}%
	\BibitemOpen
	\href@noop {} {\bibinfo  {journal} {Supplemental Material}\ }\BibitemShut
	{NoStop}%
	\bibitem [{\citenamefont {Yao}\ \emph {et~al.}(2004)\citenamefont {Yao},
		\citenamefont {Kleinman}, \citenamefont {MacDonald}, \citenamefont {Sinova},
		\citenamefont {Jungwirth}, \citenamefont {Wang}, \citenamefont {Wang},\ and\
		\citenamefont {Niu}}]{PhysRevLett.92.037204}%
	\BibitemOpen
	\bibfield  {journal} {  }\bibfield  {author} {\bibinfo {author} {\bibfnamefont
			{Y.}~\bibnamefont {Yao}}, \bibinfo {author} {\bibfnamefont {L.}~\bibnamefont
			{Kleinman}}, \bibinfo {author} {\bibfnamefont {A.~H.}\ \bibnamefont
			{MacDonald}}, \bibinfo {author} {\bibfnamefont {J.}~\bibnamefont {Sinova}},
		\bibinfo {author} {\bibfnamefont {T.}~\bibnamefont {Jungwirth}}, \bibinfo
		{author} {\bibfnamefont {D.-s.}\ \bibnamefont {Wang}}, \bibinfo {author}
		{\bibfnamefont {E.}~\bibnamefont {Wang}},\ and\ \bibinfo {author}
		{\bibfnamefont {Q.}~\bibnamefont {Niu}},\ }\href
	{https://doi.org/10.1103/PhysRevLett.92.037204} {\bibfield  {journal}
		{\bibinfo  {journal} {Phys. Rev. Lett.}\ }\textbf {\bibinfo {volume} {92}},\
		\bibinfo {pages} {037204} (\bibinfo {year} {2004})}\BibitemShut {NoStop}%
	\bibitem [{\citenamefont {Wang}\ \emph {et~al.}(2006)\citenamefont {Wang},
		\citenamefont {Yates}, \citenamefont {Souza},\ and\ \citenamefont
		{Vanderbilt}}]{PhysRevB.74.195118}%
	\BibitemOpen
	\bibfield  {author} {\bibinfo {author} {\bibfnamefont {X.}~\bibnamefont
			{Wang}}, \bibinfo {author} {\bibfnamefont {J.~R.}\ \bibnamefont {Yates}},
		\bibinfo {author} {\bibfnamefont {I.}~\bibnamefont {Souza}},\ and\ \bibinfo
		{author} {\bibfnamefont {D.}~\bibnamefont {Vanderbilt}},\ }\href
	{https://doi.org/10.1103/PhysRevB.74.195118} {\bibfield  {journal} {\bibinfo
			{journal} {Phys. Rev. B}\ }\textbf {\bibinfo {volume} {74}},\ \bibinfo
		{pages} {195118} (\bibinfo {year} {2006})}\BibitemShut {NoStop}%
	\bibitem [{\citenamefont {Qiao}\ \emph {et~al.}(2018)\citenamefont {Qiao},
		\citenamefont {Zhou}, \citenamefont {Yuan},\ and\ \citenamefont
		{Zhao}}]{PhysRevB.98.214402}%
	\BibitemOpen
	\bibfield  {author} {\bibinfo {author} {\bibfnamefont {J.}~\bibnamefont
			{Qiao}}, \bibinfo {author} {\bibfnamefont {J.}~\bibnamefont {Zhou}}, \bibinfo
		{author} {\bibfnamefont {Z.}~\bibnamefont {Yuan}},\ and\ \bibinfo {author}
		{\bibfnamefont {W.}~\bibnamefont {Zhao}},\ }\href
	{https://doi.org/10.1103/PhysRevB.98.214402} {\bibfield  {journal} {\bibinfo
			{journal} {Phys. Rev. B}\ }\textbf {\bibinfo {volume} {98}},\ \bibinfo
		{pages} {214402} (\bibinfo {year} {2018})}\BibitemShut {NoStop}%
	\bibitem [{\citenamefont {Zhou}\ \emph
		{et~al.}(2019{\natexlab{a}})\citenamefont {Zhou}, \citenamefont {Qiao},
		\citenamefont {Bournel},\ and\ \citenamefont {Zhao}}]{PhysRevB.99.060408}%
	\BibitemOpen
	\bibfield  {author} {\bibinfo {author} {\bibfnamefont {J.}~\bibnamefont
			{Zhou}}, \bibinfo {author} {\bibfnamefont {J.}~\bibnamefont {Qiao}}, \bibinfo
		{author} {\bibfnamefont {A.}~\bibnamefont {Bournel}},\ and\ \bibinfo {author}
		{\bibfnamefont {W.}~\bibnamefont {Zhao}},\ }\href
	{https://doi.org/10.1103/PhysRevB.99.060408} {\bibfield  {journal} {\bibinfo
			{journal} {Phys. Rev. B}\ }\textbf {\bibinfo {volume} {99}},\ \bibinfo
		{pages} {060408(R)} (\bibinfo {year} {2019}{\natexlab{a}})}\BibitemShut
	{NoStop}%
	\bibitem [{\citenamefont {Roman}\ \emph {et~al.}(2009)\citenamefont {Roman},
		\citenamefont {Mokrousov},\ and\ \citenamefont
		{Souza}}]{PhysRevLett.103.097203}%
	\BibitemOpen
	\bibfield  {author} {\bibinfo {author} {\bibfnamefont {E.}~\bibnamefont
			{Roman}}, \bibinfo {author} {\bibfnamefont {Y.}~\bibnamefont {Mokrousov}},\
		and\ \bibinfo {author} {\bibfnamefont {I.}~\bibnamefont {Souza}},\ }\href
	{https://doi.org/10.1103/PhysRevLett.103.097203} {\bibfield  {journal}
		{\bibinfo  {journal} {Phys. Rev. Lett.}\ }\textbf {\bibinfo {volume} {103}},\
		\bibinfo {pages} {097203} (\bibinfo {year} {2009})}\BibitemShut {NoStop}%
	\bibitem [{\citenamefont {Birss}(1964)}]{1322762}%
	\BibitemOpen
	\bibfield  {author} {\bibinfo {author} {\bibfnamefont {R.~R.}\ \bibnamefont
			{Birss}},\ }\href@noop {} {\emph {\bibinfo {title} {Symmetry and
				Magnetism}}}\ (\bibinfo  {publisher} {North-Holland Publishing Company},\
	\bibinfo {year} {1964})\BibitemShut {NoStop}%
	\bibitem [{\citenamefont {SciPy}()}]{scipy}%
	\BibitemOpen
	\bibfield  {author} {\bibinfo {author} {\bibnamefont {SciPy}},\ }\href@noop
	{} {}\bibinfo {howpublished} {\url{https://www.scipy.org/}}\BibitemShut
	{NoStop}%
	\bibitem [{\citenamefont {Yang}\ \emph {et~al.}(2021)\citenamefont {Yang},
		\citenamefont {Zhou}, \citenamefont {Feng},\ and\ \citenamefont
		{Yao}}]{PhysRevB.103.024436}%
	\BibitemOpen
	\bibfield  {author} {\bibinfo {author} {\bibfnamefont {X.}~\bibnamefont
			{Yang}}, \bibinfo {author} {\bibfnamefont {X.}~\bibnamefont {Zhou}}, \bibinfo
		{author} {\bibfnamefont {W.}~\bibnamefont {Feng}},\ and\ \bibinfo {author}
		{\bibfnamefont {Y.}~\bibnamefont {Yao}},\ }\href
	{https://doi.org/10.1103/PhysRevB.103.024436} {\bibfield  {journal} {\bibinfo
			{journal} {Phys. Rev. B}\ }\textbf {\bibinfo {volume} {103}},\ \bibinfo
		{pages} {024436} (\bibinfo {year} {2021})}\BibitemShut {NoStop}%
	\bibitem [{\citenamefont {Feng}\ \emph {et~al.}(2012)\citenamefont {Feng},
		\citenamefont {Yao}, \citenamefont {Zhu}, \citenamefont {Zhou}, \citenamefont
		{Yao},\ and\ \citenamefont {Xiao}}]{PhysRevB.86.165108}%
	\BibitemOpen
	\bibfield  {author} {\bibinfo {author} {\bibfnamefont {W.}~\bibnamefont
			{Feng}}, \bibinfo {author} {\bibfnamefont {Y.}~\bibnamefont {Yao}}, \bibinfo
		{author} {\bibfnamefont {W.}~\bibnamefont {Zhu}}, \bibinfo {author}
		{\bibfnamefont {J.}~\bibnamefont {Zhou}}, \bibinfo {author} {\bibfnamefont
			{W.}~\bibnamefont {Yao}},\ and\ \bibinfo {author} {\bibfnamefont
			{D.}~\bibnamefont {Xiao}},\ }\href
	{https://doi.org/10.1103/PhysRevB.86.165108} {\bibfield  {journal} {\bibinfo
			{journal} {Phys. Rev. B}\ }\textbf {\bibinfo {volume} {86}},\ \bibinfo
		{pages} {165108} (\bibinfo {year} {2012})}\BibitemShut {NoStop}%
	\bibitem [{\citenamefont {Zhou}\ \emph
		{et~al.}(2019{\natexlab{b}})\citenamefont {Zhou}, \citenamefont {Hanke},
		\citenamefont {Feng}, \citenamefont {Li}, \citenamefont {Guo}, \citenamefont
		{Yao}, \citenamefont {Bl\"ugel},\ and\ \citenamefont
		{Mokrousov}}]{PhysRevB.99.104428}%
	\BibitemOpen
	\bibfield  {author} {\bibinfo {author} {\bibfnamefont {X.}~\bibnamefont
			{Zhou}}, \bibinfo {author} {\bibfnamefont {J.-P.}\ \bibnamefont {Hanke}},
		\bibinfo {author} {\bibfnamefont {W.}~\bibnamefont {Feng}}, \bibinfo {author}
		{\bibfnamefont {F.}~\bibnamefont {Li}}, \bibinfo {author} {\bibfnamefont
			{G.-Y.}\ \bibnamefont {Guo}}, \bibinfo {author} {\bibfnamefont
			{Y.}~\bibnamefont {Yao}}, \bibinfo {author} {\bibfnamefont {S.}~\bibnamefont
			{Bl\"ugel}},\ and\ \bibinfo {author} {\bibfnamefont {Y.}~\bibnamefont
			{Mokrousov}},\ }\href {https://doi.org/10.1103/PhysRevB.99.104428} {\bibfield
		{journal} {\bibinfo  {journal} {Phys. Rev. B}\ }\textbf {\bibinfo {volume}
			{99}},\ \bibinfo {pages} {104428} (\bibinfo {year}
		{2019}{\natexlab{b}})}\BibitemShut {NoStop}%
	\bibitem [{\citenamefont {Wang}\ \emph {et~al.}(2007)\citenamefont {Wang},
		\citenamefont {Vanderbilt}, \citenamefont {Yates},\ and\ \citenamefont
		{Souza}}]{PhysRevB.76.195109}%
	\BibitemOpen
	\bibfield  {author} {\bibinfo {author} {\bibfnamefont {X.}~\bibnamefont
			{Wang}}, \bibinfo {author} {\bibfnamefont {D.}~\bibnamefont {Vanderbilt}},
		\bibinfo {author} {\bibfnamefont {J.~R.}\ \bibnamefont {Yates}},\ and\
		\bibinfo {author} {\bibfnamefont {I.}~\bibnamefont {Souza}},\ }\href
	{https://doi.org/10.1103/PhysRevB.76.195109} {\bibfield  {journal} {\bibinfo
			{journal} {Phys. Rev. B}\ }\textbf {\bibinfo {volume} {76}},\ \bibinfo
		{pages} {195109} (\bibinfo {year} {2007})}\BibitemShut {NoStop}%
	\bibitem [{\citenamefont {Guo}\ \emph {et~al.}(2008)\citenamefont {Guo},
		\citenamefont {Murakami}, \citenamefont {Chen},\ and\ \citenamefont
		{Nagaosa}}]{PhysRevLett.100.096401}%
	\BibitemOpen
	\bibfield  {author} {\bibinfo {author} {\bibfnamefont {G.~Y.}\ \bibnamefont
			{Guo}}, \bibinfo {author} {\bibfnamefont {S.}~\bibnamefont {Murakami}},
		\bibinfo {author} {\bibfnamefont {T.-W.}\ \bibnamefont {Chen}},\ and\
		\bibinfo {author} {\bibfnamefont {N.}~\bibnamefont {Nagaosa}},\ }\href
	{https://doi.org/10.1103/PhysRevLett.100.096401} {\bibfield  {journal}
		{\bibinfo  {journal} {Phys. Rev. Lett.}\ }\textbf {\bibinfo {volume} {100}},\
		\bibinfo {pages} {096401} (\bibinfo {year} {2008})}\BibitemShut {NoStop}%
\end{thebibliography}

%

\end{document}